\documentclass[12pt,a4paper]{article}
\usepackage[british]{babel}
\usepackage{epsfig}
\usepackage{amsmath}
\begin{document}
\date{}

\title{
{\vspace{-20mm} \normalsize
\hfill \parbox[t]{50mm}{\small DESY 07-076}}     \\[25mm]
Monte Carlo Methods in Quantum Field Theory~\footnote{
Lectures given at Spring School on High Energy Physics, 
Jaca, Spain, May 6-18, 2007}                \\[5mm]}
\author{
I. Montvay                                  \\[0.3em]
Deutsches Elektronen-Synchrotron DESY       \\[3em]}
%
\newcommand{\be}{\begin{equation}}
\newcommand{\ee}{\end{equation}}
\newcommand{\bea}{\begin{eqnarray}}
\newcommand{\eea}{\end{eqnarray}}
\newcommand{\half}{\frac{1}{2}}
\newcommand{\rar}{\rightarrow}
\newcommand{\lar}{\leftarrow}
\newcommand{\LCB}{\raisebox{-0.3ex}{\mbox{\LARGE$\left\{\right.$}}}
\newcommand{\RCB}{\raisebox{-0.3ex}{\mbox{\LARGE$\left.\right\}$}}}
\newcommand{\LSB}{\raisebox{-0.3ex}{\mbox{\LARGE$\left[\right.$}}}
\newcommand{\RSB}{\raisebox{-0.3ex}{\mbox{\LARGE$\left.\right]$}}}
\newcommand{\tr}{{\rm Tr}}
\newcommand{\I}{\ensuremath{\mathrm{i}\,}}
\newcommand{\E}{\ensuremath{\mathrm{e}\,}}

\maketitle

\abstract{In these lecture notes some applications of Monte Carlo
 integration methods in Quantum Field Theory -- in particular in
 Quantum Chromodynamics -- are introduced and discussed.}

\newpage
\section{Introduction}\label{sec1}

 The mathematical description of the Standard Model -- the theory of
 elementary particle interactions -- is based on relativistic
 Quantum Field Theory (QFT).
 Relativistic QFT is the quantum mechanics of fields defined on the
 four-dimensional space-time continuum.
 As such it has an infinite number of degrees of freedom -- the values
 of field variables in every space-time point.
 In order to define it, one has to start with the quantum theory of
 a finite number of degrees of freedom: the values of field variables
 in a finite set of discrete points within a finite volume.
 In most cases the points are lattice sites of a regular, hypercubical
 lattice over a four-dimensional torus.
 In order to define the theory one has to perform the {\em continuum
 limit} and {\em infinite volume limit} when the spacing of the lattice
 points goes to zero and the extensions of the torus grow to infinity.

 An important simplification from the mathematical point of view is
 to consider, instead of the real time variable, the time to be pure
 imaginary.
 In this {\em Euclidean space-time} the symmetry with respect to
 Lorentz-transformations becomes equivalent to the compact symmetry of
 four-dimensional rotations and, perhaps even more importantly, the
 quantum mechanical Schr\"odinger equation is transformed into an
 equation equivalent to the equation describing heat conduction
 (or e.g.~the Brownian motion).
 The consequence is that QFT with imaginary time is equivalent to
 the (classical) statistical physics of the fields.
 In the Feynman path integral formulation of quantum mechanics
 the exponent in the Boltzmann-factor is the {\em Euclidean lattice
 action}.
 (Note that the ``path'' in case of the fields is better named as the
 ``history'' of the fields in the space-time points.)

 The definition of QFT on a Euclidean space-time lattice provides a
 {\em non-perturbative regularization} without the infinities which
 have to be dealt with in perturbation theory by the renormalization
 procedure.
 One can also define perturbation theory on the lattice and in this
 way the lattice gives an alternative regularization for perturbation
 theory: the momentum cutoff is implemented by the absence of
 arbitrarily high momentum modes on the lattice.

 The number of discrete points to be considered tends to infinity
 both in the continuum limit and infinite volume limit.
 In order to differentiate between these two infinite limits one has to
 consider the ratio of the effective size of physical excitations
 to the lattice spacing.
 Obviously, this ratio has to diverge in the continuum limit.
 In the infinite volume limit, on the other hand, the ratio of the size
 of physical excitations to the volume extensions is relevant.
 In any case, one has to know about the size of the physical excitations
 which is determined by the (bare) parameters in the lattice action.
 In the language of statistical physics, in the continuum limit one has
 to tune the parameters of the lattice action to some {\em fixed point}
 with infinite correlation lengths.
 If such a fixed point exists, our knowledge in statistical physics
 suggests {\em universality}, which means that one can reach the same
 fixed point (i.e. the same continuum limit) with many different
 lattice actions.

 The most prominent example of relativistic QFT is Quantum
 Chromodynamics (QCD) which is the theory of {\em strong interactions}
 among the six known ``flavors'' of quarks: $u$-, $d$-, $s$-, $c$-
 $b$- and $t$-quark.
 QCD is a mathematically closed theory which has an unprecedented
 predictivity: it has only six independent parameters, the quark
 masses.
 More precisely the parameters of QCD are: $m_u/\Lambda_{QCD}$,
 $m_d/\Lambda_{QCD}$, $m_s/\Lambda_{QCD}$, $m_c/\Lambda_{QCD}$,
 $m_b/\Lambda_{QCD}$ and $m_t/\Lambda_{QCD}$ where the
 {\em $\Lambda$-parameter of QCD} $\Lambda_{QCD}$ is an arbitrary
 scale parameter of dimension mass.
 In many applications of QCD only the three ``light'' quarks, the
 $u$-, $d$- and $s$-quarks are relevant, therefore there are only
 three (small) parameters: $m_{u,d,s}/\Lambda_{QCD}$.
 All the properties of strong interactions as masses, decay widths,
 scattering cross-sections etc. are, in principle, determined by
 these parameters.

 The somewhat unfortunate circumstance is that, even if in principle
 determined by a very small number of free parameters, it is difficult
 to tell what are precisely the predictions of QCD.
 The reason is that strong interactions are obviously (at least
 sometimes) strong and therefore calculational methods based on
 symmetries and on perturbation theory only have a limited range of
 applicability.
 The only known method to evaluate the non-perturbative predictions of
 QCD theory is {\em lattice QCD}.
 One can formulate this in a different way by saying that the
 validation of QCD as a true theory of strong interactions is the task
 of lattice QCD theorists.

 In this series of (five) lectures on Monte Carlo methods first the
 different lattice formulations of QCD are reviewed
 (Section~\ref{sec2}).
 The basic Monte Carlo integration methods are introduced in
 Section~\ref{sec3} and discussed in some detail, including the
 important methods applicable for quark dynamics (``un-quenching'').
 Section~\ref{sec4} contains a selection of some recent developments
 in order to illustrate recent trends in lattice QCD.
 Finally, the last Section~\ref{sec5} gives a short outlook.

\section{Lattice actions}\label{sec2}

\begin{figure}[t]
\vspace*{0.01\vsize}
\begin{center}
\begin{minipage}[c]{1.0\linewidth}
\hspace{0.20\hsize}
\includegraphics[angle=-90,width=0.80\hsize]
 {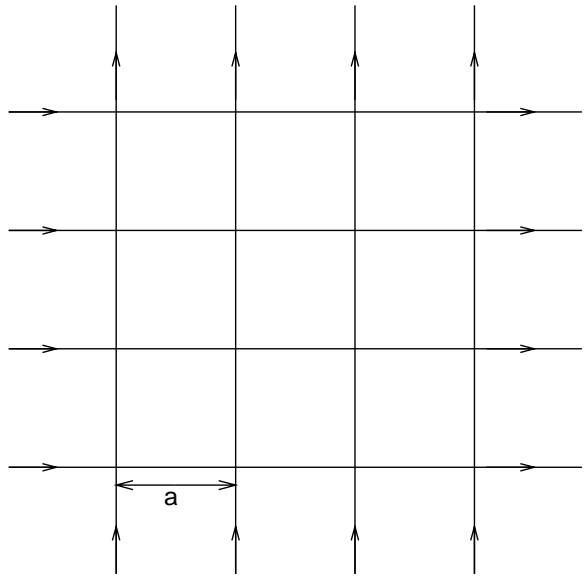}
\end{minipage}
\end{center}
\vspace*{-2em}
\begin{center}
\parbox{0.8\linewidth}{\caption{\label{fig01}\em
 A two-dimensional periodic $4 \times 4$ lattice.}}
\end{center}
\vspace*{-1em}
\end{figure}

\begin{figure}[hb]
\vspace*{-0.15\vsize}
\begin{center}
\begin{minipage}[c]{0.8\linewidth}
\hspace{0.15\hsize}
\includegraphics[angle=-90,width=0.90\hsize]
 {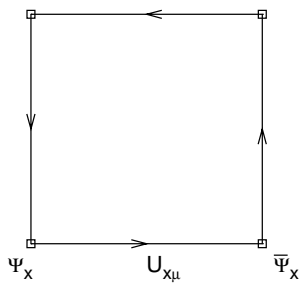}
\end{minipage}
\end{center}
\vspace*{-7em}
\begin{center}
\parbox{0.8\linewidth}{\caption{\label{fig02}\em
 The plaquette.}}
\end{center}
\vspace*{-1em}
\end{figure}

 The QFT's on the lattice are defined by their Euclidean lattice
 action.
 The lattice is in most cases a regular, hypercubical one with periodic
 boundary conditions (torus).
 Lattice elements are the {\em sites} (points) and the {\em links}
 connecting neighboring sites.
 A simple case is illustrated by the two-dimensional $4 \times 4$
 lattice in Figure~\ref{fig01}.
 The lattice spacing is usually denoted by $a$.
 For the definition of lattice gauge theories like QCD the
 {\em plaquettes} consisting of a closed path of four links are
 important (see Figure~\ref{fig02}).

 The elementary excitations in QCD are the {\em gluons} and
 {\em quarks}.
 The gluons are described by a gauge field with elements in
 the $SU(3)$ color group $U_{x\mu} \in SU(3)_{color}$ associated with
 the links $(x \to x+\hat{\mu})$ where $\hat{\mu}$ denotes the unit
 vector in the direction $\mu\;(=1,2,3,4)$.
 These are parallel transporters of the color quantum number.
 The corresponding $SU(3)$ Lie algebra element $A_{x\mu}$
 can be defined by the relation $U_{x\mu} = \exp(-aA_{x\mu})$
 with the lattice spacing $a$, in order to display the mass dimension
 of $A_{x\mu}$.
 The components of $A_{x\mu}$ are introduced by
 $A_{x\mu} = -ig A^b_\mu(x) \half\lambda_b$, with the Gell-Mann
 matrices $\lambda_b, (b=1,\ldots,8)$ and $g$ denoting the bare
 gauge coupling.
 The quark fields $\Psi$ and $\overline{\Psi}$ are associated with
 the lattice sites, as shown in Figure~\ref{fig02}.
 (For notation conventions see, in general, the book \cite{MM}.)

\subsection{Lattice actions for gluons and quarks}\label{sec2.1}

\subsubsection{The plaquette lattice action of the gauge field}
\label{sec2.1.1}

 As stated in the introduction, the lattice action for a given theory
 is not unique.
 There are large varieties of lattice actions in the same
 {\em universality class} realizing in the continuum limit the same QFT.
 For the lattice action of the $SU(3)$ color gauge field in QCD the
 simplest choice is the {\em Wilson plaquette action} introduced by Ken
 Wilson in his seminal paper on confinement and lattice QCD
 \cite{Wilson}.
 It is based on the definition of the {\em field strength}
 $F_{\mu\nu}(x)$ associated with the plaquette variable
\be\label{eq2.01}
U_{x;\mu\nu} \equiv
U_{x,\nu}^\dagger U_{x+\hat{\nu},\mu}^\dagger
U_{x+\hat{\mu},\nu}U_{x,\mu} =
\exp[-a^2 G_{\mu\nu}(x)] \ ,
\ee
 where
\be\label{eq2.02}
G_{\mu\nu}(x) = F_{\mu\nu}(x) + {\cal O}(a)
\ee
 and
\be\label{eq2.03}
F_{\mu\nu}(x) = \Delta^f_\mu A_\nu(x) - \Delta^f_\nu A_\mu(x)
+ [A_\mu(x),A_\nu(x)]
\ee
 with the {\em lattice forward derivative} defined as
 $\Delta^f\varphi(x) \equiv \varphi(x+\hat{\mu})-\varphi(x)$.

 As one can easily show, in general, for an $SU(N_c)$ color gauge
 field we have
\be\label{eq2.04}
{\rm Re\,Tr\,}U_{x;\mu\nu} = N_c +
\frac{a^4}{2} {\rm\,Tr\,}F_{\mu\nu}(x)^2 + {\cal O}(a^5)
\ee
 and therefore the Wilson (plaquette) gauge field action for the
 $SU(N_c)$ gauge field can be defined as
\bea\label{eq2.05}
S_{gauge} \equiv S_g &=& \sum_x \sum_{1 \leq\mu<\nu\leq 4} \beta 
\left\{ 1-\frac{1}{N_c}{\rm\,Re\,Tr\,}(U_{x;\mu\nu}) \right\}
\nonumber \\
&=& -\frac{\beta}{4N_c} \sum_{x\mu\nu} a^4
{\rm\,Tr\,}F_{\mu\nu}(x)F_{\mu\nu}(x) + {\cal O}(a^5) \ .
\eea
 Here we introduced the usual lattice variable for the bare gauge
 coupling as
\be\label{eq2.06}
\beta \equiv \frac{2N_c}{g^2} \ .
\ee

 An important property of the Wilson action in (\ref{eq2.05}) is
 {\em gauge invariance}.
 This is due to the fact that the trace of the product of link
 variables along any closed path is gauge invariant because the gauge
 transformation of the gauge link variables is
\be\label{eq2.07}
U^\prime_{x\mu} =
 \Lambda^{-1}(x+\hat{\mu})\;U_{x\mu}\;\Lambda(x)\hspace{2em}
 [\Lambda(x)\in SU(N_c)] \ .
\ee

 The expectation value of some function of link variables
 ${\cal O}[U]$ is given in terms of the invariant group (Haar-) measure
 $dU_{x\mu}$ as
\be\label{eq2.08}
\langle {\cal O} \rangle = \frac{1}{Z} \int\; \prod_{x\mu} dU_{x\mu}\;
\exp\{-S_{gauge}[U]\}\;{\cal O}[U] \equiv
\int[dU]\;e^{-S_{gauge}[U]}\; {\cal O}[U] \ ,
\ee
 where the {\em partition function} for the gauge field is defined as
\be\label{eq2.09}
Z = \int \prod_{x\mu} dU_{x\mu}\;\exp\{-S_{gauge}[U]\} \equiv
\int[dU]\;e^{-S_{gauge}[U]} \ .
\ee
 This shows that, indeed, in the Euclidean path integral formulation
 lattice gauge theory is equivalent to the statistical physics of
 gauge fields.

\subsubsection{The Wilson lattice action of fermion fields}
\label{sec2.1.2}

 The Dirac equation for fermions can also be similarly discretized as
 the equations of motion for the gauge field.
 A simple choice is the {\em Wilson action for fermions}:
\be\label{eq2.10}
S_q^{Wilson} = \sum_x \left\{ \mu_0 \overline{\psi}_x\psi_x
-\half\sum_\mu \overline{\psi}_{x+\hat{\mu}}\gamma_\mu U_{x\mu} \psi_x
-\frac{r}{2}\sum_\mu [\overline{\psi}_{x+\hat{\mu}}U_{x\mu}
-\overline{\psi}_x] \psi_x
\right\} \ .
\ee
 Here $\psi_x,\; \overline{\psi}_x$ are anticommuting {\em Grassmann
 variables} which have, in general, a Dirac-spinor, a color and a
 flavor index.
 For a single species (``flavor'') of fermions, of course, there is
 just a spinor and a color index.
 The lattice spacing is set now to unity: $a \equiv 1$, which is often
 done in the literature.
 $\mu_0$ is the bare quark mass in lattice units and the {\em Wilson
 parameter} is $r \ne 0$.
 The summation in (\ref{eq2.10}) runs over both positive and negative
 directions: $\sum_\mu \equiv \sum_{\mu=\pm 1}^{\pm 4}$ and, by
 definition, we have $\gamma_{-\mu} = -\gamma_\mu$.
 The role of the {\em Wilson term} proportional to $r$ will be
 discussed below.
 In (\ref{eq2.10}) the interaction of the fermion with a gauge field
 is introduced by the gauge link variables $U_{x\mu}$.
 {\em Free fermions} with no interaction correspond to $U_{x\mu}=1$.

 Often used notations are based on redefining the field normalizations
 according to
\be\label{eq2.11}
(\mu_0+4r)^{1/2}\;\psi_x \;\Rightarrow\; \psi_x \ ,  \hspace{2em}
(\mu_0+4r)^{1/2}\;\overline{\psi}_x \;\Rightarrow\; \overline{\psi}_x
\ee
 and introducing the {\em hopping parameter} by
\be\label{eq2.12}
\kappa \equiv (2\mu_0+8r)^{-1} \ , \hspace{3em}
\mu_0 = \half (\kappa^{-1}-8r) \ .
\ee
 In this way the Wilson action (\ref{eq2.10}) can be rewritten as
\be\label{eq2.13}
S_q^{Wilson} = \sum_x \left\{ (\overline{\psi}_x\psi_x) - \kappa\sum_\mu
(\overline{\psi}_{x+\hat{\mu}}U_{x\mu}[r+\gamma_\mu] \psi_x) \right\}
\equiv \sum_{xy} (\overline{\psi}_y Q_{yx} \psi_x) \ .
\ee
 In the second form the {\em Wilson fermion matrix} is (without explicit
 color- and Dirac-indices):
\be\label{eq2.14}
Q_{yx} = \delta_{yx} - \kappa\sum_\mu \delta_{y,x+\hat{\mu}}
\;U_{x\mu}\; (r+\gamma_\mu) \ .
\ee

 The particle excitations of Wilson lattice fermions can be identified
 by considering the {\em Wilson fermion propagator}, which is defined by
 the inverse of the (free) fermion matrix in (\ref{eq2.14}):
\be\label{eq2.15}
\sum_y \Delta_{zy}Q_{yx} = \delta_{zx} \ , \hspace{3em}
\Delta_{yx} = \Delta_{y-x} = \frac{1}{\Omega} \sum_k
e^{ik \cdot (y-x)} \tilde{\Delta}_k \ .
\ee
 Here $\Omega=N_1N_2N_3N_4$ is the number of lattice points and the
 allowed values of the momenta for periodic and antiperiodic boundary
 conditions, respectively, are
\be\label{eq2.16}
ap_\mu \equiv k_\mu = \frac{2\pi}{N_\mu}\nu_\mu \ ,
\hspace{2em}
k_\mu = \frac{2\pi}{N_\mu} \left(\nu_\mu+\half\right) \hspace{2em}
(\nu_\mu \in \{0,1,2,\dots,N_\mu-1\}) \ .
\ee
 Using the notations
\be\label{eq2.17}
\hat{k}_\mu \equiv 2\sin\frac{k_\mu}{2} \ , \hspace{3em}
\bar{k}_\mu \equiv \sin k_\mu \ ,
\ee
 the solution of Eq.~(\ref{eq2.15}) is given by
\be\label{eq2.18}
\tilde{\Delta}_k =
\frac{1-r\kappa(8-\hat{k}^2)-2i\kappa\gamma \cdot \bar{k}}
{[1-r\kappa(8-\hat{k}^2)]^2 + 4\kappa^2 \bar{k}^2} =
(2\kappa)^{-1}\; \frac{\mu_0+(r/2)\hat{k}^2 - i\gamma\cdot\bar{k}}
{[\mu_0+(r/2)\hat{k}^2]^2 + \bar{k}^2} \ .
\ee

 Particle excitations belong to the poles of the propagator.
 Considering the Wilson fermion propagator in (\ref{eq2.18}), it becomes
 clear why the non-zero value of the Wilson parameter $r$ is required,
 namely, for avoiding additional particle poles at $k_\mu=\pi$
 besides the physical ones at $k_\mu=0$.
 For $r=0$, which corresponds to the {\em naive discretization} of the
 Dirac equation, these additional particles emerge and -- instead of a
 single fermion flavor -- sixteen flavors are described.
 The 15 extra unphysical particles are the consequence of the first
 order character of the Dirac equation.
 Introducing a non-zero $r$ removes the unphysical fermions from the
 spectrum in the continuum limit ($a \to 0$) because their masses tend
 to infinity as $a^{-1}$.
 The price to pay for repairing the particle content is, however,
 rather high because for $r \ne 0$ the chiral symmetry is broken also
 for zero fermion mass!

\subsubsection{The Kogut-Susskind staggered lattice action of fermion
 fields}\label{sec2.1.3}

 As discussed in the previous subsection, the ``naive'' fermion action
 without the Wilson term (i.e.~$r=0$) describes 16 fermion ``flavors''.
 The {\em naive fermion action} is:
\be\label{eq2.19}
S_q^{naive} = \sum_x \left\{ \mu_0 \overline{\Psi}_x\Psi_x
+\half\sum_{\mu=1}^4 \left[
\overline{\Psi}_x\gamma_\mu \Psi_{x+\hat{\mu}}
- \overline{\Psi}_{x+\hat{\mu}}\gamma_\mu \Psi_x \right]\right\} \ .
\ee
 One can perform on this a {\em spin diagonalization} by a
 transformation
\be\label{eq2.20}
\Psi_x=A_x\psi_x \ , \hspace{4em}
\overline{\Psi}_x=\overline{\Psi}_x A^\dagger_x
\ee
 in such a way that
\be\label{eq2.21}
A^\dagger_x \gamma_\mu A_x = \alpha_{x\mu} {\bf\;1_4}
= (-1)^{x_1+\cdots+x_{\mu-1}} {\bf\;1_4} \ , \hspace{2em}
 (\mu=1,2,3,4) \ .
\ee
 One out of four identical components gives the {\em ``staggered''
 fermion action}:
\be\label{eq2.22}
S_q^{staggered} = \sum_x \left\{ \mu_0 \overline{\psi}_x\psi_x
+\half\sum_{\mu=1}^4 \alpha_{x\mu}\;\left[
\overline{\psi}_x\psi_{x+\hat{\mu}}
- \overline{\psi}_{x+\hat{\mu}}\psi_x \right]\right\} \ .
\ee

 The staggered fermion action describes four degenerate flavors with
 components scattered on the points of $2^4$ hypercubes.
 (Note that there are no Dirac spinor indices for staggered lattice
 fermions -- only color indices!)
 Rather remarkably, at zero fermion mass $\mu_0=0$ there is a remainder
 of exact chiral symmetry, namely, $\;U_{even}(1) \otimes U_{odd}(1)$.

\subsection{Improved fermion actions}\label{sec2.2}

 The freedom of choosing the lattice action in the universality class of
 the same limiting theory in the continuum can be used for:
\begin{itemize}
\item
 accelerating the convergence to the continuum limit,
\item
 achieving enhanced symmetries already at non-zero lattice spacings.
\end{itemize}
 In QCD particularly interesting is the improvement of chiral symmetry
 at non-zero lattice spacings which implies, for instance, simpler
 renormalization patterns for composite (e.g. current-) operators.

 The basic tools for constructing improved actions are {\em lattice
 perturbation theory}, {\em renormalization group transformations}
 \cite{WilsonKogut} and the {\em local effective theories} at non-zero
 cut-off \cite{Symanzik,WeiszWohlert}.

 Great effort has been invested recently in constructing {\em improved
 actions for staggered quarks} (see, for instance, the papers of the
 MILC Collaboration \cite{MILC}).
 In the so called {\em Asqtad action} the gauge action includes a
 combination of the plaquette, the $1 \times 2$ rectangle and a bent
 parallelogram 6-link term.
 The quark action includes paths up to seven links of the form
 $\overline{\psi}_y U_{y \leftarrow x} \psi_x$ where
 $U_{y \leftarrow x}$ is the product of links along the path
 $x \rightarrow y$.
 The relative weight of the contributions is such that the flavor
 symmetry breaking is suppressed and the small momentum behavior
 is improved.
 Since one staggered quark field describes four ``flavors'' of
 fermions (called here ``tastes''), for describing a single quark
 flavor in the path integral the fourth root of the fermion matrix is
 taken ({\em ``rooting''}):
\be\label{eq2.23}
\int [dU\,d\overline{\psi}\,d\psi]\,e^{-S_g-S_q} =
\int [dU]\,e^{-S_g}\det Q \;\;\Rightarrow\;\;
\int [dU]\,e^{-S_g}\,(\det Q)^{1/4} \ .
\ee
It is assumed (but debated) that this gives the correct continuum
 limit.

\subsubsection{Twisted-mass lattice QCD}\label{sec2.2.1}

 A particularly simple way of improving the Wilson-fermion action is
 the chiral rotation of the Wilson term in $S_q^{Wilson}$
 Eq.~(\ref{eq2.10}) \cite{Frezzotti:tmqcd,FrezzottiRossi:split}.
 For two equal mass quark flavors ($N_f=2$) the unbroken $SU(2)$
 subgroup of the $SU(2) \otimes SU(2)$ chiral symmetry can be partly
 rotated to axialvector directions.
 In addition, {\em ``automatic'' ${\cal O}(a)$ improvement} is
 possible \cite{FrezzottiRossi:Oa}.

 The {\em twisted mass lattice fermion action} is:
\bea\label{eq2.24}
S_q^{tm} &=& \sum_x \left\{ a\mu_q \overline{\psi}_x\psi_x
-\half\sum_\mu \overline{\psi}_{x+\hat{\mu}}\gamma_\mu U_{x\mu} \psi_x
\right. \nonumber \\
&+& \left. a\mu_{cr}\overline{\psi}_x e^{-i\omega\gamma_5\tau_3}\psi_x
-\frac{r}{2}\sum_\mu [\overline{\psi}_{x+\hat{\mu}}U_{x\mu}
-\overline{\psi}_x] e^{-i\omega\gamma_5\tau_3}\psi_x
\right\} \ .
\eea
 Here $\omega$ is the {\em twist angle}, $a\mu_q$ the bare quark mass
 in lattice units and $a\mu_{cr}=(\half\kappa_{cr}^{-1}-4r) < 0$ the
 critical bare quark mass where $\mu_q^{physical}=0$.

 The ``twist'' can be moved to the mass term by a chiral transformation
\be\label{eq2.25}
\chi_x = \exp(-\frac{i}{2}\omega\gamma_5\tau_3)\psi_x \ , \hspace{2em}
\overline{\chi}_x = 
\overline{\psi}_x \exp(-\frac{i}{2}\omega\gamma_5\tau_3) \ ,
\ee
 hence the name ``twisted mass''.
 Introducing the quark mass variables
\be\label{eq2.26}
\mu_\kappa \equiv a\mu_{cr}+a\mu_q \cos\omega \equiv \frac{1}{2\kappa}
= am_0+4r \ , \hspace{2em}
a\mu \equiv a\mu_q \sin\omega \ ,
\ee
 the action in (\ref{eq2.25}) becomes
\bea\label{eq2.27}
S_q^{tm} & = & \sum_x \left\{
\left( \overline{\chi}_x [\mu_\kappa + i\gamma_5\tau_3\,a\mu]
\chi_x \right) - \half\sum_{\mu=\pm 1}^{\pm 4}
\left( \overline{\chi}_{x+\hat{\mu}}U_{x\mu}[r+\gamma_\mu]
\chi_x \right) \right\}
\nonumber\\[0.5em]
& \equiv & \sum_{x,y} \overline{\chi}_x Q^{(\chi)}_{xy}
\chi_y \ .
\eea
 In numerical simulations one starts with this form because it
 does not contain the critical quark mass $a\mu_{cr}$ which is
 {\`a} priori unknown and has to be first numerically determined.
 Near {\em maximal twist} corresponding to $\omega=\pi/2$ it
 is also convenient to introduce till another fermion field by
 the transformations:
\be\label{eq2.28}
\Psi_x \equiv \frac{1}{\sqrt{2}} \left(1+i\gamma_5\tau_3\right)
\chi_x \ , \hspace*{3em}
\overline{\Psi}_x \equiv \overline{\chi}_x
\frac{1}{\sqrt{2}} \left(1+i\gamma_5\tau_3\right) \ .
\ee
 The {\em quark matrix on the $\chi$-basis} $Q^{(\chi)}$ defined
 in (\ref{eq2.27}) is
\be\label{eq2.29}
Q^{(\chi)}_{xy} = \delta_{xy}
\left( \mu_\kappa + i\gamma_5\tau_3\,a\mu \right)
-\half \sum_{\mu=\pm 1}^{\pm 4} \delta_{x,y+\hat{\mu}} U_{y\mu}
[r+\gamma_\mu]
\ee
 or in a short notation, without the site indices,
\be\label{eq2.30}
Q^{(\chi)} = \mu_\kappa + i\gamma_5\tau_3\,a\mu + N + R \ ,
\ee
 with
\be\label{eq2.31}
N_{xy} \equiv -\half \sum_{\mu=\pm 1}^{\pm 4} \delta_{x,y+\hat{\mu}}
U_{y\mu}\gamma_\mu \ , \hspace*{2em}
R_{xy} \equiv -\frac{r}{2} \sum_{\mu=\pm 1}^{\pm 4}
\delta_{x,y+\hat{\mu}} U_{y\mu} \ .
\ee
 On the {\em $\Psi$-basis} defined in (\ref{eq2.28}) we have the
 quark matrix
\be\label{eq2.32}
Q^{(\Psi)} = \half \left(1-i\gamma_5\tau_3\right) Q^{(\chi)}
\left(1-i\gamma_5\tau_3\right) =
a\mu + N -i\gamma_5\tau_3 \left(\mu_\kappa+R\right) \ .
\ee

 The {\em quark determinant} in the path integral over the gauge field
 is, for instance, using the quark mass variables in (\ref{eq2.24}):
\be\label{eq2.33}
\det\left[ (D^{cr}+a\mu_q\cos\omega)^\dagger (D^{cr}+a\mu_q\cos\omega)
+ (a\mu_q)^2\sin^2\omega \right]
\ee
 where the single-flavor critical fermion matrix is
\be\label{eq2.34}
D^{cr}_{yx} = a\mu_{cr}\delta_{yx}
-\half\sum_\mu \left[ \delta_{y,x+\hat{\mu}}\gamma_\mu U_{x\mu}
+r(\delta_{y,x+\hat{\mu}}U_{x\mu}-\delta_{yx})\right]
\ee

 An important feature of the twisted mass formulation is that the
 fermion matrix
\be\label{eq2.35}
D^{cr}+a\mu_q(\cos\omega+i\gamma_5\tau_3\sin\omega)
\ee
 cannot have zero eigenvalues for non-zero quark mass if
 $\omega \ne 0,\pi$.
 There are no spurious zero modes and hence {\em no exceptional gauge
 configurations} with anomalously small eigenvalues of the fermion
 matrix.
 This makes the Monte Carlo simulations at small quark- (and pion-)
 mass easier.

 The consequence of the chiral rotation corresponding to the twist is
 that the directions of vector- and axialvector-symmetries in the
 $SU(2) \otimes SU(2)$ chiral group are also rotated.
 One can achieve conserved axialvector currents but then some of the
 vector- (flavor-) symmetries will be broken.
 (The twist also induces a breaking of parity.)
 The status and consequences of the chiral symmetry can be deduced from
 the {\em chiral $SU(2) \otimes SU(2)$ Ward-Takahashi-identities}.

 Exactly conserved axialvector currents can be achieved at
 $\omega=\half\pi$.
 In this special case the conserved currents are: two axialvector
 currents ($j=1,2$ )
\bea\label{eq2.36}
A^{con}_{jx\mu} &=& \half \left\{
\left(\overline{\psi}_{x+\hat{\mu}}\gamma_\mu\gamma_5\frac{\tau_j}{2}
U_{x\mu}\psi_x \right)
+ \left(\overline{\psi}_x\gamma_\mu\gamma_5\frac{\tau_j}{2}
U_{x\mu}^\dagger\psi_{x+\hat{\mu}} \right)
\right. \nonumber \\
&+& \left.
r\left(\overline{\psi}_{x+\hat{\mu}}\frac{\overline{\tau}_j}{2}
U_{x\mu}\psi_x \right)
- r\left(\overline{\psi}_x \frac{\overline{\tau}_j}{2}
U_{x\mu}^\dagger\psi_{x+\hat{\mu}} \right)\right\}
\eea
 with $\overline{\tau}_1 \equiv \tau_2$ and
 $\overline{\tau}_2 \equiv -\tau_1$, and one vector current:
\bea\label{eq2.37}
V^{con}_{3x\mu} &=& \half \left\{
\left(\overline{\psi}_{x+\hat{\mu}}\gamma_\mu\frac{\tau_3}{2}
U_{x\mu}\psi_x \right)
+ \left(\overline{\psi}_x\gamma_\mu\frac{\tau_3}{2}
U_{x\mu}^\dagger\psi_{x+\hat{\mu}} \right)
\right. \nonumber \\
&-& \left.
\frac{ir}{2}\left(\overline{\psi}_{x+\hat{\mu}}\gamma_5
U_{x\mu}\psi_x \right)
+\frac{ir}{2}\left(\overline{\psi}_x \gamma_5
U_{x\mu}^\dagger\psi_{x+\hat{\mu}} \right)\right\} \ .
\eea

 The invariance of the path integral with respect to the change of
 variables
\bea\label{eq2.38}
\psi^\prime_x &=& (1+\frac{i}{2}\alpha_{Vrx}\tau_r
 +\frac{i}{2}\alpha_{Arx}\gamma_5 \tau_r)\psi_x \ ,
\nonumber \\
\overline{\psi}^\prime_x &=& \overline{\psi}_x
(1-\frac{i}{2}\alpha_{Vrx}\tau_r +
\frac{i}{2}\alpha_{Arx}\gamma_5\tau_r)
\eea
 implies for an arbitrary function ${\cal O}$ of field variables the
 following WT-identities:
\bea\label{eq2.39}
\left\langle {\cal O}\;\Delta^b_\mu A^{con}_{jx\mu} \right\rangle
+ \left\langle\frac{{\cal O}\stackrel{\leftarrow}{\partial}}
{\partial\psi_x}\; \gamma_5\frac{\tau_j}{2}\psi_x
+ \overline{\psi}_x\gamma_5\frac{\tau_j}{2}\;
\frac{\stackrel{\rightarrow}{\partial}{\cal O}}
{\partial\overline{\psi}_x} \right\rangle
&=& \mu_q \left\langle {\cal O}\;\overline{\psi}_x\gamma_5\tau_j\psi_x
\right\rangle
\nonumber \\
\left\langle {\cal O}\;\Delta^b_\mu V^{con}_{3x\mu} \right\rangle
+ \left\langle\frac{{\cal O}\stackrel{\leftarrow}{\partial}}
{\partial\psi_x} \;\frac{\tau_3}{2}\psi_x
- \overline{\psi}_x\frac{\tau_3}{2}\;
\frac{\stackrel{\rightarrow}{\partial}{\cal O}}
{\partial\overline{\psi}_x} \right\rangle &=& 0 \ ,
\eea
 with the backward lattice derivative defined as
 $\Delta^b_\mu\varphi(x) \equiv \varphi(x)-\varphi(x-\hat{\mu})$.

 Besides the conserved axialvector currents the important feature
 of twisted-mass Wilson fermions is {\em automatic ${\cal O}(a)$
 improvement.}
 (${\cal O}(a)$ improvement means that in the continuum limit
 $a\to 0$ the leading deviation from the limiting value behaves
 asymptotically as ${\cal O}(a^2)$.)
 As it has been shown by Frezzotti and Rossi \cite{FrezzottiRossi:Oa},
 for the (untwisted) Wilson fermion action we have
\be\label{eq2.40}
\langle{\cal O}\rangle^{WA}_{(m_q)} \equiv
\half\left[ \langle{\cal O}\rangle_{(r,m_q)}
+\langle{\cal O}\rangle_{(-r,m_q)} \right] \propto
\langle{\cal O}\rangle^{cont}_{(m_q)} + {\cal O}(a^2) \ .
\ee
 This is averaging over opposite sign Wilson parameters:
 ``Wilson average''.

 In twisted mass lattice QCD (tmLQCD) changing the sign of $r$ is
 equivalent to shifting the twist angle by $\pi$.
 In the special case of $\omega=\half\pi$ this is equivalent to
 $\omega \rightarrow -\omega$, therefore expectation values even in
 $\omega$ are ``automatically'' ${\cal O}(a)$ improved, without any
 averaging.
 Automatically ${\cal O}(a)$ improved physical quantities are, for
 instance:
\begin{itemize}
\item
 the energy eigenvalues, hence the masses;
\item
 on-shell matrix elements at zero spatial momenta;
\item
 matrix elements of operators with parity equal to the product of the
 parities of the external states.
\end{itemize}

\subsubsection{Domain wall lattice fermions}\label{sec2.2.2}

 The chiral symmetry of massless fermions can be realized at non-zero
 lattice spacing by introducing a fifth ``extra dimension''
 \cite{Kaplan,NarayananNeuberger,Shamir}.
 In the fifth direction there is a ``defect'': either the mass term
 changes sign \cite{Kaplan} or there are ``walls'' at the two ends
 \cite{Shamir}.
 In this case there are chiral fermion solutions which are exponentially
 localized in the fifth dimension near these defects.
 The gauge field remains four-dimensional (independent on the fifth
 dimension).
 In the limit of infinitely large fifth dimension the positive and
 negative chirality solutions (at opposite walls or at opposite sign
 changes on a torus) have zero overlap with each other and the chiral
 symmetry becomes exact.

 The domain wall fermion action can be written (with
 $1 \leq s \leq N_s$) as
\be\label{eq2.41}
S_F = \sum_{s,s^\prime} \overline{\Psi}_{xs}
(D_F)_{x s , x^\prime s^\prime} \Psi_{x^\prime s^\prime}
\ee
 where in an $s$-block form
\be\label{eq2.42}
D_F = \left(
\begin{array}{cccccccc}
\sigma+D    & -\sigma P_L &  0          &  0          & \ldots &
0           & 0           &  m_f P_R    \\[0.7em]
-\sigma P_R & \sigma+D    & -\sigma P_L &  0          & \ldots &
0           & 0           &  0          \\[0.7em]
0           & -\sigma P_R & \sigma+D    & -\sigma P_L & \ldots &
0           & 0           &  0          \\[0.7em]
\vdots      & \vdots      & \vdots      & \vdots      & \ddots &
\vdots      & \vdots      & \vdots      \\[0.7em]
0           & 0           & 0           &  0          & \ldots &
-\sigma P_R & \sigma+D    & -\sigma P_L \\[0.7em]
m_f P_L     & 0           & 0           &  0          & \ldots &
0           & -\sigma P_R & \sigma+D    \end{array}   \right) \ .
\ee
 The chiral projectors are denoted, as usual, by
 $P_{R,L} \equiv \half (1 \pm \gamma_5)$, the quark mass in lattice
 units is $m_f$, the ratio of lattice spacings is $\sigma = a/a_s$
 and the four-dimensional Wilson-Dirac matrix with negative mass
 ($0 > -m_0 > -2$) is, for $r=1$,
\be\label{eq2.43}
D_{x x^\prime} = (4-m_0) \delta_{x x^\prime}\;
-\; \half \sum_{\mu=1}^4 \left[
\delta_{x^\prime,x+\hat{\mu}}(1+\gamma_\mu) U_{x\mu}
+ \delta_{x^\prime+\hat{\mu},x}(1-\gamma_\mu) U^\dagger_{x^\prime\mu}
 \right] \ .
\ee

 The hermitian fermion matrix corresponding to $D_F$ in (\ref{eq2.42})
 is useful, for instance, in Monte Carlo simulations.
 It can be constructed as follows: since with an $s$-reflection
 $(R_5)_{s s^\prime} \equiv \delta_{N_s+1-s,s^\prime}$ we have
\be\label{eq2.44}
D_F = R_5\gamma_5 D_F^\dagger R_5\gamma_5 \ ,
\ee
 the hermitian fermion matrix can be defined as
\be\label{eq2.45}
\tilde{D}_F \equiv R_5\gamma_5 D_F = \tilde{D}_F^\dagger \ .
\ee

 The chiral symmetry is broken by a non-zero overlap of the opposite
 chirality wave functions, which tends to zero in the limit of an
 infinite extension of the fifth dimension: $N_s\to\infty$.
 Enhanced symmetry breaking occurs if the four-dimensional Wilson
 fermion matrix $D$ has small eigenvalues.

\subsubsection{Neuberger overlap fermions}\label{sec2.2.3}

 Another possibility to achieve chiral symmetry of the lattice fermion
 action, which in fact can be related to domain wall lattice
 fermions, is the {\em Neuberger (overlap-) fermion action.}

 Let us rewrite the (free) Wilson fermion action for $r=1$ and
 $\mu_0 \equiv am_0$ as
\bea\label{eq2.46}
S_q^{Wilson} &=& \sum_x a^4\, \overline{\psi}_x [ m_0 + D_W ] \psi_x \ ,
\nonumber \\
D_W &\equiv& \sum_{\mu=1}^4 \left[\half\gamma_\mu(\nabla_\mu+\nabla^*_\mu)
-\frac{a}{2}\, \nabla^*_\mu\nabla_\mu \right] \ ,
\eea
 where the lattice derivatives are now denoted by
\be\label{eq2.47}
\nabla_\mu \equiv a^{-1}\Delta^f_\mu \ , \hspace{3em}
\nabla^*_\mu \equiv a^{-1}\Delta^b_\mu \ .
\ee
 The Neuberger lattice fermion operator with zero mass is defined as
\be\label{eq2.48}
D_N \equiv 
\frac{1}{a}\left(1-A\,\frac{1}{\sqrt{A^\dagger A}}\right) \ , 
\hspace{3em}  A \equiv 1 - aD_W \ .
\ee
 The inverse square-root here can be realized by polynomial or rational
 approximations.
 Note that $A$ is proportional to the Wilson fermion matrix with bare
 mass $-a^{-1}$.

 An important property of the Neuberger operator $D_N$ is that
 $V \equiv 1-aD_N$ is unitary: $V^\dagger V = 1$.
 As a consequence, the spectrum of $D_N=a^{-1}(1-V)$ is on a circle
 going through the origin.
 In addition, the Neuberger operator satisfies the {\em Ginsparg-Wilson
 relation}
\be\label{eq2.49}
\gamma_5 D_N + D_N \gamma_5 = aD_N \gamma_5 D_N \ .
\ee
 This is equivalent to the condition as introduced by Ginsparg and Wilson
 (GW) \cite{GinspargWilson}
\be\label{eq2.50}
\gamma_5 D^{-1}+ D^{-1} \gamma_5 = 2aR\gamma_5 \ .
\ee
 The GW-relation is the optimal approximation to chiral symmetry which
 can be realized by a lattice fermion operator for $a \to 0$.
 $R$ in (\ref{eq2.50}) is, in general, a local operator.
 For the Neuberger operator $D=D_N$ we have $R=\half$.

 The {\em lattice chiral symmetry} satisfied by a GW-lattice fermion
 can be explicitely displayed by appropriately defined chiral
 transformations \cite{Luscher:chiral}.
 It can be shown that
\be\label{eq2.51}
\delta\psi = \gamma_5 \left(1-\frac{a}{2}D \right)\psi \ , \hspace{3em}
\delta\overline{\psi} =
\overline{\psi}\left(1-\frac{a}{2}D \right)\gamma_5
\ee
 is an exact chiral symmetry for any lattice spacing $a$ if the
 GW-relation is satisfied.

 Lattice actions satisfying the GW-relation are:
\begin{itemize}
\item
 the fixed point action, which is the fixed point of some
 renormalization group transformation \cite{Hasenfratz};
\item
 the Neuberger action $D_N$ in (\ref{eq2.48});
\item
 the effective (four-dimensional) action of the light fermion field of
 the domain wall fermion \cite{NeubergerKikukawa}.
\end{itemize}

 Note: the inverse of the effective Dirac operator of the light fermion
 field of the domain wall fermion is equivalent to the inverse of the
 truncated overlap Dirac operator (except for a local contact term).
 Using GW-fermions one can prove the index theorem about topological
 charge \cite{HasenfratzLalienaNiedermayer} and introduce the
 $\theta$-parameter in QCD, etc.

 Having lattice actions with exact chiral symmetry at non-zero lattice
 spacing is a great achievement.
 Although it is expected that (spontaneously broken) chiral symmetry
 is restored in the continuum limit also for simple lattice
 formulations with, for instance, Wilson fermions, the explicit
 breaking of chiral symmetry for non-zero lattice spacings makes
 the renormalization of composite operators more involved and in practice
 also much more cumbersome because of the extended mixing pattern.
 The chiral symmetry restricts the mixing to be simpler and more
 tractable.
 
 The difficulty of defining chiral symmetric lattice actions is
 emphasized by the {\em Nielsen-Ninomiya theorem}
 \cite{NielsenNinomiya}.
 This theorem states that there is no (free) lattice fermion action
 which can be written in the form
\be\label{eq2.52}
S_f = a^4 \sum_{xy} \overline{\psi}_y D(y-x)\psi_x
\ee
 and which would simultaneously satisfy the following conditions:
\begin{itemize}
\item
 $D(x)$ is local (bounded for large $x$ by $e^{-\gamma|x|}$),
\item
 its Fourier-transform is
 $\tilde{D}(p) = i\gamma_\mu p_\mu+{\cal O}(ap^2)$ for $p \ll \pi/a$,
\item
 $\tilde{D}(p)$ is invertible for $p \ne 0$ (i.e. there are no massless
 fermion doubler poles),
\item
 $\gamma_5 D + D \gamma_5 = 0$ (chiral symmetry).
\end{itemize}

 GW-fermions circumvent the Nielsen-Ninomiya theorem by relaxing the
 last condition: instead of exact anticommutativity only a weaker
 condition, namely the Ginsparg-Wilson relation in (\ref{eq2.49}), is
 satisfied.
 Correspondingly, the chiral transformation is modified: the simple
 continuum transformation is generalized to (\ref{eq2.51}).

 The important question is whether the locality of the action is
 ensured for GW-fermions.
 In case of the Neuberger (overlap) action locality can be proven if
 the gauge field is smooth enough, namely if every plaquette value is
 close to unity \cite{HernandezJansenLuscher}.
 Because of the importance of locality such gauge fields are sometimes
 called {\em ``admissible''}.
 Of course, usual lattice actions typically admit any plaquette value
 and therefore in the path integral ``inadmissible'' configurations
 also occur.
 In fact, in actual simulations there are always plaquettes with small
 values.
 It is an open question whether this turns out to be a problem in the
 continuum limit.
 In any case, the lattice spacing has to be small enough in order
 to avoid the {\em ``Aoki phase''} with lots of small eigenvalues of
 $D_W$.
 The small eigenvalues make $D_N$ non-local and the ``residual mass''
 breaking the chiral symmetry of domain wall fermions large
 \cite{GoltermanShamir:aoki}.

\section{Monte Carlo integration methods}\label{sec3}

 The goal of numerical simulations in Quantum Field Theories (QFT's) is
 to estimate the expectation value of some functions $A[\varphi]$ of
 the field variables generically denoted by
 $[\varphi] \equiv \{ \varphi_{x\alpha} \}$.
 In terms of path integrals this is given as
\be\label{eq3.53}
\langle A \rangle = Z^{-1} \int [d\varphi]
e^{-S[\varphi]} A[\varphi]  \ ,
\hspace{3em}
Z = \int [d\varphi] e^{-S[\varphi]} \ .
\ee
 $S[\varphi]$ is the lattice action, which is assumed to be
 a real function of the field variables.
 (To begin with, we only consider bosonic path integrals.)

 A typical lattice action contains a summation over the lattice sites.
 Since the number of lattice points $\Omega$ is large, there are many
 integration variables.
 However, since (\ref{eq3.53}) corresponds to a statistical system
 with a large number of degrees of freedom, in the path integral only a
 small vicinity of the minimum of the ``free energy'' density will
 substantially contribute.
 A suitable mathematical method to treat with such situations is
 {\em Monte Carlo integration}.
 (For a recent review of Monte Carlo integration in QFT's see
 Ref.~\cite{Morningstar}.)

\subsection{Monte Carlo integration}\label{sec3.1}

\subsubsection{Simple Monte Carlo integration}\label{sec3.1.1}

 Let us consider a continuous real function $f(X)$ of a continuous
 random variable $X$ having probability distribution $p_X(s)$ and hence
 the {\em expectation value}
\be\label{eq3.54}
\langle f(X) \rangle = \int ds\, f(s)\, p_X(s) \ .
\ee
 Using $p_X(s)$ to obtain $N$ outcomes of $X$  ($X_1,X_2,\ldots,X_N$), the
 random variables $Y_j=f(X_j)$ give
\be\label{eq3.55}
\lim_{N\to\infty}\frac{1}{N}\sum_{j=1}^N Y_j = \langle Y \rangle =
\langle f(X) \rangle = \int ds\, f(s)\, p_X(s) \ .
\ee
 In a short notation:
\be\label{eq3.56}
\overline{f} \equiv \frac{1}{N}\sum_{j=1}^N f(X_j) , \hspace{3em}
\lim_{N\to\infty} \overline{f} = \langle f \rangle =
\int ds\, f(s)\, p_X(s) \ .
\ee

 For large $N$, the central limit theorem tells us that the error in
 approximating $\langle f(X) \rangle$ is given by the variance
 $V[f(X)]$ as $\sqrt{V[f(X)]/N}$.
 The Monte Carlo estimate of the variance is:
\be\label{eq3.57}
V[Y] = \left\langle (Y-\langle Y \rangle)^2 \right\rangle \approx
\overline{(f-\overline{f})^2} = \overline{f^2} - \overline{f}^2 \ .
\ee
 Generalizing this to several ($D$) integration variables one obtains
 the following formulas for simple Monte Carlo integration:
\be\label{eq3.58}
\int_{\cal V} d^Dx\, p(\vec{x})\, f(\vec{x}) \approx
\overline{f} \pm
\left( \frac{\overline{f^2} - \overline{f}^2}{N} \right)^\half \ .
\ee
 Here, according to the notation introduced in (\ref{eq3.56}),
\be\label{eq3.59}
\overline{f} \equiv \frac{1}{N}\sum_{i=1}^N f(\vec{x}_i) ,
\hspace*{3em}
\overline{f^2} \equiv \frac{1}{N}\sum_{i=1}^N f(\vec{x}_i)^2 \ .
\ee
 The points $\vec{x}_1,\vec{x}_2,\ldots,\vec{x}_N$ have to be
 chosen {\em independently} and {\em randomly} with probability
 distribution $p(\vec{x})$ in the $D$-dimensional volume $\cal V$.

\subsubsection{Importance sampling}\label{sec3.1.2}

 Simple Monte Carlo integration works best for flat functions but is
 problematic if the integrand is sharply peaked or rapidly oscillating.
 Therefore, a good procedure is to apply {\em importance sampling}:
 find a positive function $g(x)$ with integral norm unity
 ($\int dx\, g(x) = 1$) such that $h(x) \equiv f(x)/g(x)$ is as close
 as possible to a constant and then calculate
\be\label{eq3.60}
\int_a^b dx\,f(x) = \int_a^b dx\,g(x)h(x) \approx
\frac{(b-a)}{N}\sum_{j=1}^N h(x_j) \ ,
\ee
 where the points $x_j$ are chosen with probability density $g(x)$ and
 we used simple Monte Carlo integration with a constant probability in
 an interval:
\be\label{eq3.61}
\int_a^b dx\,f(x) \approx \frac{(b-a)}{N}\sum_{j=1}^N f(x_j) \ .
\ee
 The prerequisite is, of course, that one can find an appropriate
 $g(x)$ such that on can generate points with it.

 How can one generate the desired (in general, multi-dimensional)
 probability distributions?
 One possibility for lower-dimensional integrals is the {\em rejection
 method}.
 This is based on the observation that sampling with $p_X(x)$, for
 instance, in an interval $x \in [b,a]$ is equivalent to choose a
 random point uniformly in two dimensions and reject it unless it is
 in the area under the curve $p_X(x)$.
 For high-dimensional distributions this becomes cumbersome.
 Multi-dimensional integrals can be handled by exploiting {\em Markov
 processes}.

\subsubsection{Markov chains}\label{sec3.1.3}

 A Markov process (or {\em ``Markov chain''}) is a sequence of states
 which are generated with {\em transition probabilities} from a given
 state to the next one.
 The transition probability is assumed to depend only on the current
 state of the system and not on any previous state.
 For simplicity, for discrete states $s_1,s_2,\ldots,s_R$ the
 transition probability can be denoted by $p_{ij}$.
 The matrix $\bf P$ with elements $p_{ij}$ is called {\em transition
 matrix} (or {\em Markov-matrix}).

 The mathematical properties of Markov chains are extensively covered
 in the literature.
 For a comprehensive collection of features relevant in Monte Carlo
 integration of QFT's see Ref.~\cite{Morningstar}. 
 Let us mention here just a few of them:
\begin{itemize}
\item
 The product of two Markov matrices $\bf P_1 P_2$ is again a Markov
 matrix.
\item
 Every eigenvalue of a Markov matrix satisfies $|\lambda| \leq 1$.
\item
 Every Markov matrix has at least one eigenvalue $\lambda=1$.
\end{itemize}

 A very important statement is given by the  {\em fundamental limit
 theorem} for (irreducible, aperiodic) Markov chains: they have a
 unique stationary distribution satisfying
 ${\bf w}^T = {\bf w}^T {\bf P}$ which is identical to the limiting
 distribution $w_j = \lim_{n\to\infty} p^{(n)}_{ij}$.

 An important concept is the {\em autocorrelation} in Markov chains.
 Since the state of the system depends on the  previous state, the
 consecutive states are not uncorrelated.
 To reach a more or less uncorrelated distribution from some initial
 one, in general, several steps have to be performed.
 The degree of correlation among the subsequent states can be
 characterized by the {\em autocorrelation function} which is defined
 for some observable $O_i$ as
\be\label{eq3.62}
\rho(t) \equiv
\left( \langle O_i O_{i+t} \rangle - \langle O_i \rangle^2 \right)
\left/\left( \langle O_i^2 \rangle - \langle O_i \rangle^2 \right.\right)
\ .
\ee
 Obviously, decreasing autocorrelations decrease the Monte Carlo error
 for a given length of the Markov chain.

\subsection{Updating}\label{sec3.2}

 The aim in Monte Carlo simulations of QFT's is to calculate the
 expectation values of some functions of field variables as given in
 (\ref{eq3.53}).
 The Monte Carlo integration is based on {\em importance sampling}.
 The required distribution of field configurations according to
 the {\em Boltzmann factor} $e^{-S[\varphi]}$ ({\em ``canonical
 distribution}``) is generated by a Markov chain by exploiting the
 {\em fundamental limit theorem} discussed in Section~\ref{sec3.1.3}.

 Let us denote the configuration sequence generated in the Markov
 chain by $\{ [\varphi_n],\; 1 \le n \le N \}$.
 In this {\em field configuration sample} the expectation values are
 approximated by the sample average:
\be\label{eq3.63}
\overline{A} \equiv \frac{1}{N} \sum_{n=1}^N A[\varphi_n]\hspace{1em}
\stackrel{N\to\infty}{\Longrightarrow}\hspace{1em}
\langle A \rangle \ .
\ee

 The Markov process of generating one field configuration after the
 other is generally called {\em updating}.
 Let us denote the transition probability from a configration to
 the next one $[\varphi] \to [\varphi^{\prime}]$ by
 $P\left( [\varphi^{\prime}] \lar [\varphi] \right)$.
 In order to generate the canonical distribution $e^{-S[\varphi]}$ a
 sufficient condition is
\be\label{eq3.64}
P\left( [\varphi^{\prime}] \lar [\varphi] \right) e^{-S[\varphi]} =
P\left( [\varphi] \lar [\varphi^{\prime}] \right)
e^{-S[\varphi^{\prime}]}\ .
\ee
 This condition is called {\em detailed balance}.

\subsubsection{Metropolis algorithm}\label{sec3.2.1}

 The ``ancestor'' of updating processes for bosonic systems is
 the Metropolis algorithm \cite{Metropolis}.
 For a system with ${\cal N}$ possible configurations the transition
 probability for $[\varphi^{\prime}] \neq [\varphi]$ is defined by
\be\label{eq3.65}
P([\varphi^{\prime}] \lar [\varphi]) = {\cal N}^{-1}\;
\min\left\{ 1, \frac{e^{-S[\varphi^{\prime}]}}{e^{-S[\varphi]}} 
\right\} \ .
\ee
 This transition matrix can be realized by the following numerical
 procedure:
\begin{quote}
 i.) choose first a trial configuration randomly from ${\cal N}$
 configurations and
 ii.) accept it as the next configuration in any case if the Boltzmann
 factor is increased (the action is decreased).
 If the Boltzmann factor is decreased (the action is increased), then
 accept the change with probability equal to the ratio of the Boltzmann
 factors.
\end{quote}
 The {\em accept-reject step} can be implemented by comparing the ratio
 of the Boltzmann factors to a pseudo-random number between 0 and 1.
 One can see by inspection that the above transition probability
 distribution satisfies the detailed balance condition (\ref{eq3.64}),
 hence it creates the desired canonical distribution of configurations.

\subsubsection{Fermions in Monte Carlo simulations}\label{sec3.2.2}

 The lattice action for QFT's with fermions, for instance like QCD,
 has the generic form
\be\label{eq3.66}
S[U,\psi,\overline{\psi}] =
 S_g[U]+S_q[U,\psi,\overline{\psi}] \ ,
\ee
 where $S_g$ is the bosonic part, in QCD the color gauge field part,
 and $S_q$ is describing the fermion fields and their interaction with
 the bosonic fields.
 $S_q$ is assumed to be quadratic in the Grassmann-variables of the
 fermion fields:
\be\label{eq3.67}
S_q = \sum_{xy} (\overline{\psi}_y Q_{yx} \psi_x) \ .
\ee
 The expectation values have the general form
\be\label{eq3.68}
\langle F \rangle = \frac{ \int [d U\, d\overline{\psi}\, d\psi]
e^{-S_g-S_q} F[U,\psi,\overline{\psi}] }
{ \int [d U\, d\overline{\psi}\, d\psi] e^{-S_g-S_q} } \equiv
Z^{-1} \int [d U\, d\overline{\psi}\, d\psi]
e^{-S_g-S_q} F[U,\psi,\overline{\psi}] \ .
\ee
 After performing the Grassmann integration one obtains
\bea\label{eq3.69}
\left\langle \psi_{y_1} \overline{\psi}_{x_1}
             \psi_{y_2} \overline{\psi}_{x_2}
\cdots       \psi_{y_n} \overline{\psi}_{x_n} F[U] \right\rangle =
Z^{-1} \int [d U] e^{-S_g[U]}\; \det Q[U]\; F[U] &&
\nonumber \\
\cdot \sum_{z_1 \cdots z_n}
\epsilon^{z_1 z_2 \cdots z_n}_{y_1 y_2 \cdots y_n} \;
Q[U]^{-1}_{z_1x_1} Q[U]^{-1}_{z_2x_2} \cdots Q[U]^{-1}_{z_nx_n} \ . &&
\eea
 Here $Q[U]^{-1}$ is an (external) quark propagator and $\det Q[U]$
 generates the virtual quark loops.

 Since taking into account the {\em fermion determinant} $\det Q[U]$
 in the path integral over the bosonic (gauge-) fields is a very
 demanding computational task, in a crud approximation one sometimes
 simply omits it.
 This is called {\em ``quenched approximation''}:
 $\det Q[U]\Rightarrow 1$.
 Experience in QCD shows that the results in the quenched approximation
 are often qualitatively reasonable, nevertheless the error caused by
 omitting the closed virtual fermion loops is uncontrollable and
 implies the presence of unphysical {\em ``ghost'' contributions}.

\subsubsection{Dynamical fermions: ``unquenching''}\label{sec3.2.3}

 In the early days of lattice QCD simulations quite often the quenched
 approximation was taken.
 This is, however, on the long run not acceptable, the obtained
 results do not represent a numerical solution of QCD.
 More recently -- due to some impressive developments in the available
 computer power and in our algorithmic skills -- the true dynamical
 simulation of quarks became feasible.

 The basic difficulty in {\em ``unquenching''} is that the fermion
 determinant is a non-local function of the bosonic fields and
 therefore it is a great challenge for computations.
 For solving this problem a useful tool is the {\em pseudofermion
 representation} \cite{PetcherWeingarten}:
\be\label{eq3.70}
\det\,(Q^\dagger Q) \propto \int [d\phi\, d\phi^+] \exp\left\{
-\sum_{xy} (\phi^+_y [Q^\dagger Q]^{-1}_{yx} \phi_x) \right\} \ .
\ee
 In case of, for instance, Wilson quarks the quark determinant
 satisfies
\be\label{eq3.71}
Q^\dagger=\gamma_5 Q \gamma_5 
\hspace{2em}\Longrightarrow\hspace{2em}
\det Q^\dagger = \det Q \ ,
\ee
 therefore Eq.~(\ref{eq3.71}) describes the quark determinant of
 {\em two degenerate quark flavors}.

 In the popular {\em Hybrid Monte Carlo (HMC)} algorithm \cite{HMC}
 the representation (\ref{eq3.70}) is implemented in the updating by
 using molecular dynamics equations (see Section~\ref{sec3.3}).
 For single quark flavors HMC is not applicable.
 One can, however, use {\em Polynomial Hybrid Monte Carlo (PHMC)}
 \cite{PHMC,MontvayScholz} (see Section~\ref{sec3.4})
 or {\em Rational Hybrid Monte Carlo (RHMC)} \cite{RHMC}.

\subsection{Hybrid Monte Carlo}\label{sec3.3}

\subsubsection{HMC for gauge fields}\label{sec3.3.1}

 The basic idea of HMC is to employ {\em molecular dynamics (MD)}
 equations in order to collectively move the field configuration in the
 whole lattice volume.
 Since discretized molecular dynamics equations are used, the lattice
 action (analogous to the energy in molecular dynamics) is not conserved
 along {\em MD-trajectories}, therefore at the end of a trajectory a
 Metropolis {\em accept-reject step} has to be implemented.
 In this subsection HMC will be introduced in the important case of
 lattice gauge fields, specifically $SU(3)$ (color) gauge field.

 The equations of motion are derived from a Hamiltonian which is defined
 for the colour gauge field $U_{x,\mu} \in {\rm SU(3)}$ as
\be\label{eq3.72}
H[P,U] = \half\sum_{x\mu j} P_{x\mu j}^2 + S_g[U] \ ,
\ee
 where $S_g[U]$ is the gauge field action and the real variables
 $P_{x\mu j},\; j=1,\ldots,8$ are called  {\em conjugate momenta}.
 They are the expansion coefficients of the Lie algebra element
\be\label{eq3.73}
P_{x,\mu} \equiv \sum_j i\lambda_j P_{x\mu j} \ .
\ee
 It is assumed that the conjugate momenta have a Gaussian distribution:
\be\label{eq3.74}
P_{x\mu j} \propto \exp\left\{ -\half\sum_{x\mu j} P_{x\mu j}^2 \right\}
\equiv P_M[P] \ .
\ee
 The expectation value of some function $F[U]$ is defined as
\be\label{eq3.75}
\left\langle F \right\rangle = 
\frac{\int [d\,P][d\,U]\exp(-H[P,U])\, F[U]}
{\int [d\,P][d\,U]\exp(-H[P,U])} \ .
\ee

 By a proper choice of the discretized trajectories one can achieve
 that the transition probability from a configuration to the next
 satisfies {\em detailed balance} (see next subsection).
 Therefore, the correct canonical distribution is reproduced.

 The Hamiltonian equations of motion are:
\be\label{eq3.76}
\frac{d\,P_{x\mu j}}{d\,\tau} = -D_{x\mu j} S_g[U] \ , \hspace{2em}
\frac{d\,U_{x\mu}}{d\,\tau} = iP_{x,\mu} U_{x,\mu} \ ,
\ee
 where the derivative with respect to the gauge field is defined,
 in general, as
\be\label{eq3.77}
D_{x\mu j} f[U] \equiv \left.\frac{d}{d\,\alpha}\right|_{\alpha=0}
f\left( e^{i\alpha\lambda_j}\, U_{x,\mu} \right) \ .
\ee
%

\subsubsection{Detailed balance}\label{sec3.3.2}

 In order to prove that HMC reproduces the correct canonical
 distribution of (gauge) fields it is sufficient to prove the detailed
 balance condition (\ref{eq3.64}) for the transition probabilities
 realized by the MD-trajectories.

 The discretized trajectories $T_H$ provide the following transition
 probability distribution at the end of the trajectory:
\be\label{eq3.78}
P_H   \left( [P^{\prime},U^{\prime}] \lar [P,U] \right) =
\delta\left( [P^{\prime},U^{\prime}]  -  T_H[P,U] \right) \ .
\ee
 Let us assume that the trajectories satisfy {\em reversibility}:
\be\label{eq3.79}
P_H   \left( [P^{\prime},U^{\prime}] \lar [P,U] \right) =
P_H \left( [-P,U] \lar [-P^{\prime},U^{\prime}] \right) \ .
\ee
 The Metropolis acceptance step is described by the well known
 probability distribution
\be\label{eq3.80}
P_A \left( [P^{\prime},U^{\prime}] \lar [P,U] \right) =
\min \left\{ 1, e^{-H[P^{\prime},U^{\prime}] + H[P,U]}
\right\} \ .
\ee
 The total transition probability is then
\be\label{eq3.81}
P \left( [U^{\prime}] \lar [U] \right) =
\int [dP\, dP^{\prime}]
P_A \left( [P^{\prime},U^{\prime}] \lar [P,U] \right)
P_H \left( [P^{\prime},U^{\prime}] \lar [P,U] \right) P_M[P] \ .
\ee
 Using the relation
\be\label{eq3.82}
e^{ -H[P,U] }
\min \left\{ 1, e^{-H[P^{\prime},U^{\prime}] + H[P,U]}
\right\} = e^{ -H[P^{\prime},U^{\prime}] }
\min \left\{ 1, e^{-H[P,U] + H[P^{\prime},U^{\prime}]}
\right\} \ ,
\ee
 one shows
\bea\label{eq3.83} 
e^{ -H[P,U] } P_A \left( [P^{\prime},U^{\prime}] \lar [P,U] \right) &=&
e^{ -H[P^{\prime},U^{\prime}] } 
P_A \left( [P,U] \lar [P^{\prime},U^{\prime}] \right)
\nonumber \\
&=& e^{ -H[-P^{\prime},U^{\prime}] }
P_A \left( [-P,U] \lar [-P^\prime,U^\prime] \right) \ .
\eea
 Therefore, due to reversibility, we have for the canonical
 distribution
\be\label{eq3.84}
W_c[U] \propto \exp\,\{ -S_g[U] \}
\ee
 the relation
\bea\label{eq3.85}
&& W_c[U] \int [dP\, dP^{\prime}]
P_A \left( [P^{\prime},U^{\prime}] \lar [P,U] \right)
P_H \left( [P^{\prime},U^{\prime}] \lar [P,U] \right) P_M[P]
\\ \nonumber
&& = W_c[U^{\prime}] \int [dP\, dP^{\prime}]
P_A \left( [-P,U] \lar [-P^{\prime},U^{\prime}] \right)
 P_H \left( [-P,U] \lar [-P^{\prime},U^{\prime}] \right)
P_M[-P^{\prime}] \ .
\eea
 Taking into account that
\be\label{eq3.86}
[dP\, dP^{\prime}] = [d(-P)\, d(-P^{\prime})] \ ,
\ee
 this is just the detailed balance condition we wanted to prove.

\subsubsection{Leapfrog trajectories}\label{sec3.3.3}

 The proof of detailed balance for HMC in the previous subsection has
 been based on the assumption that the discretized MD-trajectories are
 reversible.
 The classical example is a {\em leapfrog trajectory} which is defined
 as follows.

 First we update the conjugate momente with a step size
 $\Delta\tau=\half\delta\tau$.
 This is followed by $(n-1)$ update steps with
 $\Delta\tau=\half\delta\tau$ both for the gauge variables and for
 the momentum variables, alternating with each other.
 Finally, the gauge variables are updated with $\Delta\tau=\delta\tau$
 and the momentum variables with $\Delta\tau=\half\delta\tau$.

 The explicit formulae for these steps are:
\bea\label{eq3.87}
P^\prime_{x\mu j} &=& P_{x\mu j} - D_{x\mu j}S_g[U]\,\Delta\tau
\nonumber \\
U^\prime_{x,\mu} &=&
\exp\left\{ \sum_j i\lambda_j\,P_{x\mu j}\,\Delta\tau \right\}
U_{x,\mu} \ .
\eea
 One can easily prove that the reversibility condition (\ref{eq3.79})
 is satisfied.

 The single steps in the leapfrog trajectory cause a discretization
 error of the order $\delta\tau^3$.
 Therefore, the action for the final configuration is expected to
 differ from the initial configuration by an error of order
 $\delta\tau^2$.

 In the second equation of (\ref{eq3.87}) we need, in each step on a
 trajectory for each link, the evaluation of the exponential of an
 element of the gauge group algebra $A$.
 It is desirable to minimize the cost of this, but at the same time
 the calculation has to be precise enough for not loosing
 reversibility.
 Since one can show that
\be\label{eq3.88}
A^3 = \left( \half {\rm Tr\,}A^2 \right)\, A
+ \left( \frac{1}{3} {\rm Tr\,}A^3 \right)\, I \ ,
\ee
 any analytic function $f(A)$ can be written as
\be\label{eq3.89}
f(A) = a_2\, A^2 + a_1\, A +a_0\, I \ .
\ee
 For the exponential function $f(A)=\exp(A)$ the coefficients
 $a_{0,1,2}$ can be practically calculated by recursion relations
 based on the Taylor expansion of $\exp(A)$.

\subsubsection{HMC for QCD}\label{sec3.3.4}

 Besides the color gauge field dealt with in the previous subsections,
 in QCD one has to introduce the quarks, too.
 Let us consider here two equal mass quarks, in order to be able to
 replace the fermionic quark fields by bosonic {\em pseudofermion
 fields} according to (\ref{eq3.70}).
 (Single quark flavors will be considered in the next
 Section~\ref{sec3.4}.)

 Let us note that the pseudofermion field in (\ref{eq3.70}) is an
 auxiliary complex scalar field $\phi_{qx \alpha c}$ having the same
 number of components as the fermion field $\psi_{qx \alpha c}$.
 (The indices in QCD are: $q$ for the quark flavors, $x$ for lattice
 sites, $\alpha$ for the Dirac spinor index and $c$ for color.)
 According to (\ref{eq3.70}) the fermion determinant induces an
 {\em effective action} for the gauge field which can be written as
\be\label{eq3.90}
S_{eff}[U] \equiv
\sum_{xy} (\phi^+_y \{ Q[U]^+Q[U] \}^{-1}_{yx} \phi_x) \ .
\ee
 In the MD-trajectories of the previous subsections $S_{eff}[U]$
 has to be added to the pure gauge action:
\be\label{eq3.91}
S_g[U] \Longrightarrow S_g[U]+S_{eff}[U] \ .
\ee
%

\subsection{Polynomial Hybrid Monte Carlo}\label{sec3.4}

 Here we discuss the PHMC algorithm \cite{PHMC} with multi-step
 stochastic corrections \cite{MontvayScholz}.
 This update algorithm is applicable for any number of quark flavors,
 provided that the fermion determinant is positive, which is the case
 for {\em positive quark mass}.
 For negative quark masses there is a {\em sign problem}, which will
 not be discussed here.

 For $N_f=1,2,\ldots$ degenerate quarks one uses
\be\label{eq3.92}
\left|\det(Q)\right|^{N_f} =
\left\{\det(Q^\dagger Q) \right\}^{N_f/2} =
\left\{\det(\tilde{Q}^2) \right\}^{N_f/2}
\simeq \frac{1}{\det P_n(\tilde{Q}^2)} \ ,
\ee
 where the {\em Hermitian fermion matrix} is
 $\tilde{Q} \equiv \gamma_5 Q$ and the polynomial $P_n$ satisfies
\be\label{eq3.93}
\lim_{n \to \infty} P_n(x) = x^{-N_f/2}
\ee
 in an interval $[\epsilon,\lambda]$ covering the spectrum of
 $Q^\dagger Q = \tilde{Q}^2$.

 The effective gauge action representing the fermions in the path
 integral is now
\be\label{eq3.94}
S_{eff}[U] =
\sum_{xy} (\phi^+_y P_n(\tilde{Q}^2)_{yx} \phi_x) \ .
\ee
 Sometimes it is more effective to simulate several fractional quark
 flavors:
\be\label{eq3.95}
\left( \det \tilde{Q}^2 \right)^{N_f/2} =
\left[\left( \det \tilde{Q}^2 \right)^{N_f/(2n_B)} \right]^{n_B} \ ,
\ee
 which can be called {\em determinant break-up.}
 In this case we need a polynomial approximation
\be\label{eq3.96}
\hspace*{1em} P_n(x) \simeq x^{-\alpha}
\ee
 with
\be\label{eq3.97}
\alpha \equiv \frac{N_f}{2n_B}
\ee
 and positive integer $n_B$.
 The effective gauge action with determinant break-up has then
 {\em multiple pseudofermion fields}:
\be\label{eq3.98}
S_{eff}[U] =
\sum_{k=1}^{n_B}\sum_{xy} (\phi^+_{ky} P_n(\tilde{Q}^2)_{yx} \phi_{kx})
\ .
\ee

 Since polynomial approximations with a finite $n$ cannot be exact,
 one has to correct for the committed error.
 One can show that for small fermion masses in lattice units $am \ll 1$
 the (typical) smallest eigenvalue of $\tilde{Q}^2$ behaves as $(am)^2$
 and for a fixed quality of approximation within the interval
 $[\epsilon,\lambda]$ the degree of the polynomial is growing as
\be\label{eq3.99}
n \propto \sqrt{\epsilon} \propto (am)^{-1}  .
\ee
 This would require in realistic simulations very high degree
 polynomials with $n \geq 10^3$-$10^4$.
 The way out is to perform {\em stochastic corrections} during
 the updating process \cite{MontvayScholz}.

 This goes as follows:
 for improving the approximation a second polynomial is introduced
 according to
\be\label{eq3.100}
P_1(x) P_2(x) \simeq x^{-\alpha} \ , \hspace{2em}
x \in [\epsilon,\lambda] \ .
\ee
 The first polynomial $P_1(x)$ gives a crude approximation
\be\label{eq3.101}
P_1(x) \simeq x^{-\alpha} \ .
\ee
 The second polynomial $P_2(x)$ gives a good approximation
 according to
\be\label{eq3.102}
P_2(x) \simeq [x^\alpha P_1(x)]^{-1} \ .
\ee
 (This can also be extended to a multi-step approximation
 \cite{MontvayScholz}.)

 During the updating process $P_1$ is realized by PHMC updates
 \cite{PHMC}, whereas $P_2$ is taken into account stochastically by a
 {\em noisy correction step}.
 This goes as follows: one generates a Gaussian random vector with
 distribution
\be\label{eq3.103}
\frac{e^{-\eta^\dagger P_2(\tilde{Q}[U]^2)\eta}}
{\int [d\eta] e^{-\eta^\dagger P_2(\tilde{Q}[U]^2)\eta}}
\ee
 and accepts the change $[U] \rar [U^\prime]$ with probability
\be\label{eq3.104}
\min\left\{ 1,A(\eta,[U^\prime] \lar [U]) \right\} \ ,
\ee
 where
\be\label{eq3.105}
A(\eta,[U^\prime] \lar [U]) =
\exp\left\{-\eta^\dagger P_2(\tilde{Q}[U^\prime]^2)\eta
           +\eta^\dagger P_2(\tilde{Q}[U]^2)\eta\right\} \ .
\ee
 It can be shown that this update procedure satisfies the detailed
 balance condition.

 The Gaussian noise vector $\eta$ can be obtained from $\eta^\prime$
 distributed according to the simple Gaussian distribution
\be\label{eq3.106}
\frac{e^{-\eta^{\prime\dagger}\eta^\prime}}
{\int [d\eta^\prime] e^{-\eta^{\prime\dagger}\eta^\prime}}
\ee
 by setting it equal to
\be\label{eq3.107}
\eta = P_2(\tilde{Q}[U]^2)^{-\half} \eta^\prime \ .
\ee
 In order to obtain the inverse square root on the right hand side
 one can proceed with a polynomial approximation
\be\label{eq3.108}
\bar{P}_2(x) \simeq P_2(x)^{-\half} \ , \hspace{3em}
x \in [\bar{\epsilon},\lambda] \ .
\ee
 The interval $[\bar{\epsilon},\lambda]$ can be chosen differently
 from the approximation interval $[\epsilon,\lambda]$ for $P_2$,
 usually with $\bar{\epsilon} < \epsilon$.

 The polynomial approximation with $P_2$ can only become exact in the
 limit when the degree $n_2$ of $P_2$ is infinite.
 Instead of investigating the dependence of expectation values on
 $n_2$ by performing several simulations and extrapolating to
 $n_2\to\infty$, one fixes $n_2$ to some high value and performs
 another correction in the expectation values by still finer
 polynomials.
 This is done by {\em reweighting} the configurations.
 This {\em measurement correction} is based on a further polynomial
 approximation $P^\prime$ with degree $n^\prime$ which satisfies
\be\label{eq3.109}
\lim_{n^\prime \to \infty} P_1(x)P_2(x)P^\prime(x) =
x^{-\alpha} \ , \hspace{3em}
x \in [\epsilon^\prime,\lambda] \ .
\ee
 The interval $[\epsilon^\prime,\lambda]$ can be chosen such that
 $\epsilon^\prime=0,\lambda=\lambda_{max}$, where $\lambda_{max}$ is an
 absolute upper bound of the eigenvalues of $\tilde{Q}^2$.

 In practice it is more effective to take $\epsilon^\prime > 0$ and
 determine the eigenvalues below $\epsilon^\prime$ and the
 corresponding correction factors exactly.
 For the evaluation of $P^\prime$ one can use recursive relations,
 which can be stopped by observing the required precision of the
 result.

 After reweighting the expectation value of a quantity $A$ is given by
\be\label{eq3.110}
\langle A \rangle = \frac{
\langle A \exp{\{\eta^\dagger[1-P^\prime(\tilde{Q}^2)]\eta\}}
\rangle_{U,\eta}}
{\langle  \exp{\{\eta^\dagger[1-P^\prime(\tilde{Q}^2)]\eta\}}
\rangle_{U,\eta}} \ ,
\ee
 where $\eta$ is a simple Gaussian noise.
 Here $\langle\ldots\rangle_{U,\eta}$ denotes an expectation value on
 the gauge field sequence, which is obtained in the two-step process
 described before, and {\em on a sequence of independent $\eta$'s of
 arbitrary length}.

 In most practical applications of PHMC with stochastic correction
 the second step (or the last step if multiple correction is applied)
 of the polynomial approximation can be chosen precise enough such
 that the deviation from the exact results is negligible compared to
 the statistical errors.
 In such cases the reweighting is not necessary.
 However, for very small fermion masses reweighting may become a more
 effective possibility than to choose very high order polynomials
 for a good enough approximation.

 A positive aspect of reweighting is related to the change of the
 topological charge of the gauge configurations.
 Such changes occur through configurations with zero eigenvalues
 of the fermion determinant where the molecular dynamical force
 becomes infinite.
 This implies an infinite barrier for changing the topological charge
 which may completely suppress transitions between the topological
 sectors.
 This problem is substantially weakened by PHMC algorithms because
 the polynomial approximations do not reproduce the singularity of
 the inverse fermion determinant (i.e.~ the zero of the determinant)
 \cite{FrezzottiJansen}.
 In this way the gauge configuration can tunnel between topological
 sectors.
 The more frequent occurrence of the configurations near the zeros of
 the fermion determinant is corrected by the reweighting.

\subsubsection{PHMC and twisted mass}\label{sec3.4.1}

 Until now we tacitly assumed that we use ordinary (``untwisted'')
 fermions.
 In case of twisted mass lattice QCD the numerical simulation of
 light quarks is, in fact, easier, because the quark determinant of a
 degenerate quark doublet becomes, according to Eq.~(\ref{eq2.33}),
\be\label{eq3.111}
\det{(\tilde{Q}^2+\mu_s^2)}
\ee
 where $\mu_s\equiv\mu_q\sin\omega$ with $\mu_q$ the quark mass in
 lattice units and $\omega$ the twist angle.

 The polynomials $P_{1,n_1}(x)$ and $P_{2,n_2}(x)$ now satisfy
\be\label{eq3.112}
\lim_{n_2 \to \infty} P_{1,n_1}(x)\,P_{2,n_2}(x) =
(x+\mu_s^2)^{-N_f/2} \ , \hspace{3em}
x \in [\epsilon,\lambda] \ .
\ee
 In case of $\omega\simeq\frac{\pi}{2}$ the polynomial approximations
 have lower orders and the updating is faster because of the
 {\em absence of exceptional configurations} with very small
 eigenvalues, due to the presence of the lower limit $\mu_s^2$.
 (Note that the very small eigenvalues are often originating from
 {\em topological defects} at the cutoff scale, which are unphysical
 lattice artifacts going away in the continuum limit.)

\section{Some recent developments}\label{sec4}

 In spite of substantial algorithmic developments, lattice QCD
 simulations near the small (physical) quark masses still need rather
 high computer power: we need Tflops!
 An example for a demanding Monte Carlo simulation (in the near future)
 is: $\Omega = 50^3 \cdot 100 = 1.25 \cdot 10^7$ and $am_q = 0.005$.
 This is equivalent, for instance, at $a = 0.1\,{\rm fm}$ to
 $m_q=10\,{\rm MeV},\; L=5\,{\rm fm},\; m_\pi \simeq 200\,{\rm MeV}$.

 The smallness of the $u$-, $d$- and $s$-quark masses implies that
 the numerical simulation (with dynamical quarks) is a great challenge
 for computations.
 There are a number of large international collaborations working
 on this problem over the world:
\begin{itemize}
\item
 USA: MILC, RBC, ... Collaboration;
\item
 Japan: CP-PACS, JLQCD, ... Collaboration;
\item
 Europe: UKQCD, Alpha, QCDSF, ETM ... Collaboration.
\end{itemize}

 It would be rather difficult to give a review of all the interesting
 results achieved over the last years.
 Here I shall only give a very limited and personal collection of
 some of the problems and results.

\subsection{The light pseudoscalar boson sector}\label{sec4.1}

\subsubsection{Gasser-Leutwyler coefficients}\label{sec4.1.1}

 The physical consequence of the smallness of three quark masses is the
 existence of eight light pseudo-Goldstone bosons: $\pi,K,\eta$.
 In the low-energy pseudo-Goldstone boson sector there is an
 ${\rm SU(3)}\otimes{\rm SU(3)}$ chiral flavour symmetry and the
 dynamics can be described by {\em Chiral Perturbation Theory (ChPT)}
 \cite{Weinberg,GasserLeutwyler}.
 In an expansion in powers of momenta and light quark masses several
 low energy constants -- the {\em Gasser-Leutwyler constants} --
 appear which parameterize the strength of interactions in the low
 energy chiral Lagrangian.

 An eminent task for Monte Carlo simulations in Lattice-QCD is to
 describe the pseudo-Goldstone boson sector.
 The Gasser-Leutwyler constants are free parameters which can be
 constrained by analyzing experimental data.
 In the framework of {\em lattice regularization} they can be
 determined from first principles by numerical simulations.
 In numerical simulations, besides the possibility of changing momenta,
 one can also change the masses of the quarks.

 ChPT can be extended by changing the {\em valence quark masses} in
 quark propagators independently from the {\em sea quark masses} in
 virtual quark loops.
 In this way one arrives at {\em Partially Quenched Chiral Perturbation
 Theory (PQChPT)} \cite{BernardGolterman} (see Section \ref{sec4.1.3}).

\subsubsection{E(uropean) T(wisted) M(ass) Collaboration}\label{sec4.1.2}

 This collaboration consists of about 30 physicists from 7 countries:
\begin{enumerate}
\item
 Cyprus: University of Cyprus,
\item
 France: Universit\'e de Paris Orsay,
\item
 Germany: DESY, Universit\"at M\"unster, TU M\"unchen,
\item
 Italy: Universit\`a di Roma I,II,III, INFN, ECT$^*$,
\item
 Spain: Universidad Val\`encia,
\item
 Switzerland: ETH Z\"urich,
\item
 United Kingdom: University of Liverpool.
\end{enumerate}
 In a recent paper (first of a series) numerical Monte Carlo simulations
 on {\em ``Dynamical Twisted Mass Fermions with Light Quarks''} are
 reported \cite{ETMC}.

\begin{figure}[t]
\vspace*{0.01\vsize}
\begin{minipage}[c]{0.90\linewidth}
\begin{flushleft}
\hspace{0.05\hsize}
\includegraphics[angle=0,width=0.45\hsize]
 {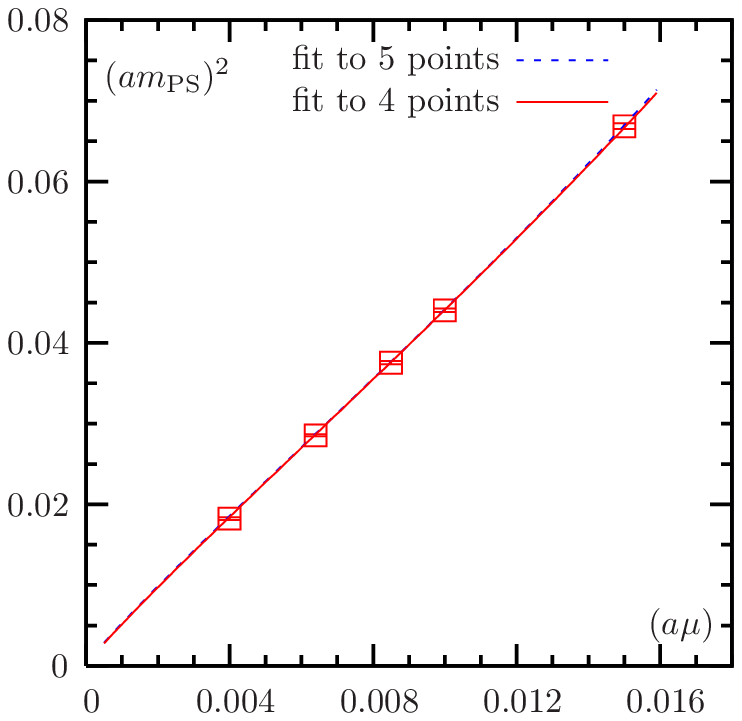}
\end{flushleft}
\vspace*{-0.331\vsize}
\begin{flushright}
\hspace{0.05\hsize}
\includegraphics[angle=0,width=0.45\hsize]
 {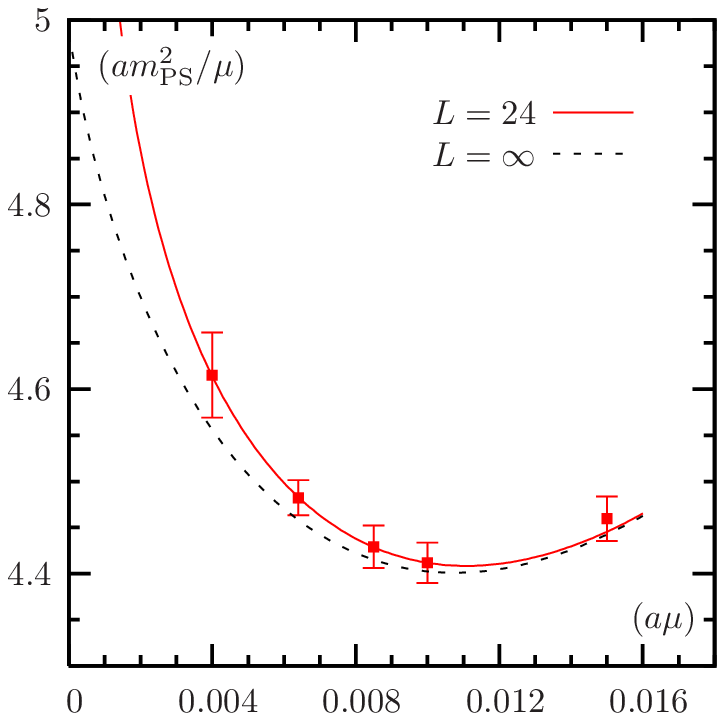}
\end{flushright}
\end{minipage}
\vspace*{-0.1em}
\begin{center}
\parbox{0.8\linewidth}{\caption{\label{fig03}\em
 Left: $(am_\pi)^2$ as a function of the twisted mass $a\mu$;
 right: $(am_\pi)^2/(a\mu)$ versus $a\mu$ (by the ETM Collaboration).
 The finite volume ChPT-fit is shown, together with the infinite
 volume limit (dashed line): $\bar{l}_3 = 3.65(12)$.}}
\end{center}
\vspace*{-1em}
\end{figure}

\begin{figure}[hbt]
\vspace*{0.01\vsize}
\begin{minipage}[c]{0.90\linewidth}
\begin{flushleft}
\hspace{0.05\hsize}
\includegraphics[angle=0,width=0.45\hsize]
 {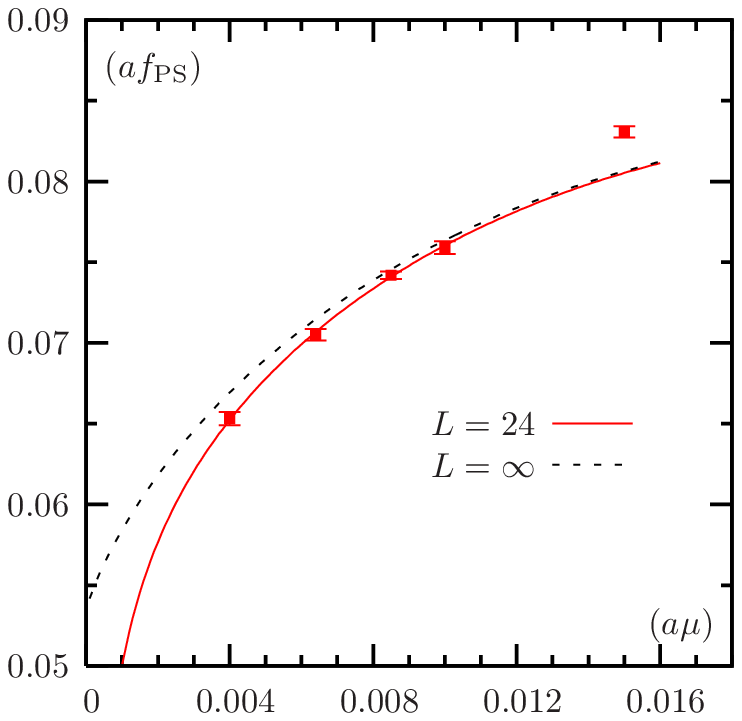}
\end{flushleft}
\vspace*{-0.330\vsize}
\begin{flushright}
\hspace{0.05\hsize}
\includegraphics[angle=0,width=0.45\hsize]
 {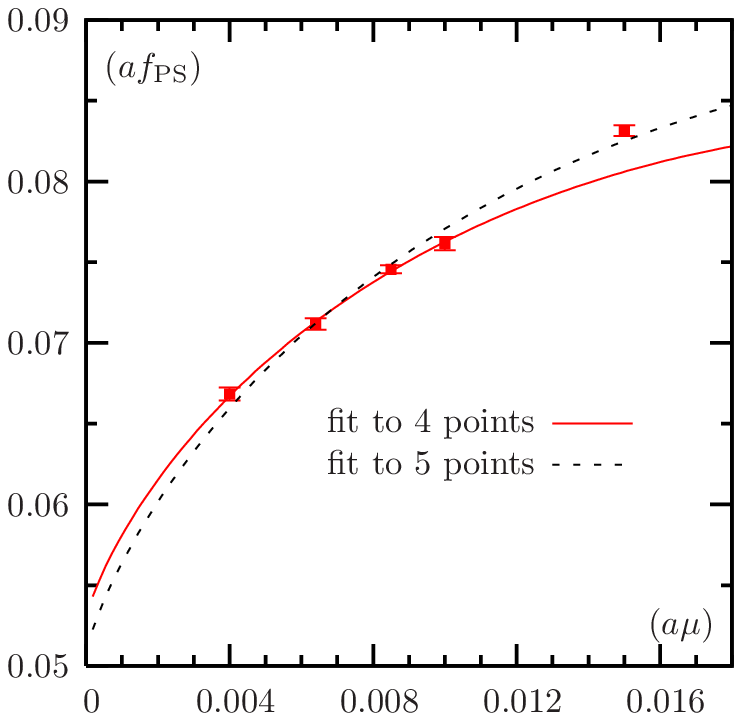}
\end{flushright}
\end{minipage}
\vspace*{-0.1em}
\begin{center}
\parbox{0.8\linewidth}{\caption{\label{fig04}\em
 ChPT fits to $af_\pi$ versus $a\mu$ (by the ETM Collaboration).
 Left: the point with largest $a\mu$ left out (the dashed line is the
 infinite volume limit);
 right: compared to finite volume fit to every point.
 The fit gives: $a = 0.087(1) {\rm\, fm}$,
 $a^{-1} = 2264(26) {\rm\, MeV}$), $\;\;\bar{l}_4 = 4.52(06)$.}}
\end{center}
\vspace*{-1em}
\end{figure}

 As examples of the results, {\em Chiral Perturbation Theory (ChPT)}
 fits of the pseudo\-sca\-lar- (pion-) mass (in Figure~\ref{fig03}) and
 pseudoscalar- (pion-) decay constant (in Figure~\ref{fig04}) are
 shown.
 It is remarkable that the precision on $\bar{l}_{3,4}$ is much higher
 than obtained by any previous experimental determination.
 However: this is with only $N_f=2$ degenerate dynamical quarks
 ($u$- and $d$-quark) and no continuum limit extrapolation is yet
 performed (it is comming soon).

\subsubsection{Ratio tests of PQChPT}\label{sec4.1.3}

 Taking ratios at fixed gauge coupling ($\beta$) is advantageous
 because the Z-factors of {\em mutiplicative renormalization} cancel
 (for instance, in $m_q$ and $f_\pi$).
 Also: some types of lattice artifacts may cancel.

 In case of simulations with Wilson-type lattice actions, by taking
 into account lattice artifacts in the Chiral Lagrangian, one can
 reach the continuum limit faster.
 This approach is based on the {\em effective continuum theory}
 introduced by Symanzik \cite{Symanzik}: cutoff effects (in the lattice
 regularized theory) can be described by ${\cal O}(a,a^2,\ldots)$ terms
 in a {\em local effective Lagrangian}.

 This idea can be applied to low energy LQCD
 \cite{SharpeSingleton,RupakShoresh}.
 In case of the Wilson quark action the leading ${\cal O}(a)$ effects
 have a simple chiral transformation property, identical to those of
 the quark masses.
 At leading order of ChPT, besides the quark mass variable $\chi$, an
 additional ${\cal O}(a)$ parameter $\rho$ appears:
\be\label{eq4.113}
\chi \equiv \frac{2B_0 m_q}{f_0^2} \ , \hspace{3em}
\rho \equiv \frac{2W_0 a}{f_0^2}       \hspace{4em}
\left( \eta \equiv \frac{\rho}{\chi} \right) \ .
\ee
 At next to leading order (NLO): the Gasser-Leutwyler constants
 $L_1,\ldots,L_8$ are doubled by the (bare parameter dependent)
 coefficients $W_1,\ldots,W_8$ describing ${\cal O}(a)$ effects.
 (Extension to ${\cal O}(a^2)$ is possible.)

 Variables to be used in {\em ratio tests of PQChPT} (the index $V$
 always stands for ``valence'' quarks which are ``quenched'',
 $S$ for dynamical ``sea'' quarks):
\be\label{eq4.114}
\xi   \equiv \frac{m_{qV}}{m_{qS}} =\frac{\chi_V}{\chi_S} \ ,
\hspace{3em}
\eta_S  \equiv \frac{\rho_S}{\chi_S} \ ,
\hspace{3em}
\sigma_i \equiv \frac{m_{qS}^{(i)}}{m_{qS}^{(R)}} =
\frac{\chi_S}{\chi_R} \ .
\ee
 For the pion decay constants the appropriate ratios are:
\be\label{eq4.115}
Rf_{VV} \equiv \frac{f_{VV}}{f_{SS}} \ ,  \hspace{3em}
Rf_{VS} \equiv \frac{f_{VS}}{f_{SS}} \ ,  \hspace{3em}
RRf \equiv \frac{f_{VS}^2}{f_{VV}f_{SS}} \ ,
\ee
 and for the pion mass-squares (dividing by the leading order
 behaviour):
\be\label{eq4.116}
Rn_{VV} \equiv \frac{m_{VV}^2}{\xi m_{SS}^2} \ ,     \hspace{2em}
Rn_{VS} \equiv \frac{2m_{VS}^2}{(\xi+1) m_{SS}^2} \ ,\hspace{2em}
RRn \equiv \frac{4\xi m_{VS}^4}{(\xi+1)^2m_{VV}^2 m_{SS}^2} \ .
\ee
 For the sea quark mass dependence
\be\label{eq4.117}
Rf_{SS} \equiv \frac{f_{SS}}{f_{RR}} \ , \hspace{4em}
Rn_{SS} \equiv \frac{m_{SS}^2}{\sigma m_{RR}^2}
\ee
 are appropriate.

 Examples of the NLO formulas are \cite{RupakShoresh,qq+q}:
 for $N_s$ degenerate sea quarks
\bea\label{eq4.118}
Rf_{VV} &=& 1 + 4(\xi-1)\chi_S L_{S5}
+ \frac{N_s\chi_S}{32\pi^2}(1+\eta_S)\log(1+\eta_S)
\nonumber \\
&& - \frac{N_s\chi_S}{64\pi^2}(1+\xi+2\eta_S)\log\frac{1+\xi+2\eta_S}{2}\ ,
\\ \label{eq4.119}
RRf &=& 1 + \frac{\chi_S}{32N_s\pi^2}(\xi-1)
- \frac{\chi_S}{32N_s\pi^2}(1+\eta_S)\log\frac{\xi+\eta_S}{1+\eta_S} \ ,
\\ \label{eq4.120}
Rf_{SS} &=& 1 + 4(\sigma-1)\chi_R (N_s L_{R4}+L_{R5})
+ 4(\eta_S\sigma-\eta_R)\chi_R (N_s W_{R4}+W_{R5})
\nonumber \\
&& - \frac{N_s\chi_R}{32\pi^2}\sigma(1+\eta_S)\log[\sigma(1+\eta_S)]
+\frac{N_s\chi_R}{32\pi^2}(1+\eta_R)\log(1+\eta_R)  \ ,
\eea
 and similarly for $Rn\ldots$.

 In the above formulas $L_{Sk}$ denote Gasser-Leutwyler constants
 renormalized at the scale $f_0\sqrt{\chi_S}$.
 They are related to $\bar{L}_k$ defined at the scale $f_0$ and
 $L^\prime_k$ defined at the generic scale $\mu$ according to
\be\label{eq4.121}
L_{Sk} = \bar{L}_k - c_k\log(\chi_S)
= L^\prime_k - c_k\log(\frac{f_0^2}{\mu^2}\chi_S) \ ,
\ee
 with some (known) constants $c_k$.
 The corresponding relations for the coefficients $W_{Sk}$ are:
\be\label{eq4.122}
W_{Sk} = \bar{W}_k - d_k\log(\chi_S)
= W^\prime_k - d_k\log(\frac{f_0^2}{\mu^2}\chi_S) \ .
\ee
 Note that these formulas can be extended to the NNLO order, too.

 A first comparison of these formulas with numerical Monte Carlo
 results has been performed by the {\em qq+q Collaboration}
 (DESY-M\"unster) \cite{qq+q}.
 The lattice sizes were $16^4$ and $16^3 \cdot 32$, and $N_s=2$ light
 quark flavours were simulated.
 The lattice spacing was:
 $a=0.189(5)\, {\rm fm} \simeq (1.04\, {\rm GeV})^{-1}\;$
 giving lattice extensions $L \simeq 3\, {\rm fm}$.
 The pion masses were:
 $am_\pi = 0.6747(14),\; 0.6211(22),\; 0.4354(68),\; 0.3676(23)$
 which correspond in physical units to
 $m_\pi \simeq 702,\; 646,\; 452,\; 415\, {\rm MeV}$.
 The sea quark masses were approximately $60\,{\rm MeV}$ to
 $25\,{\rm MeV}$; and the valence quark masses:
 $\frac{1}{2} m_{sea} \leq m_{valence} \leq 2 m_{sea}$.
 Being the first exploratory study, the parameters did not correspond
 to the latest best ones, in particular, the lattice spacing was
 rather coarse and the quark masses not small enough.

\begin{figure}[t]
\vspace*{0.01\vsize}
\begin{center}
\begin{minipage}[c]{1.0\linewidth}
\hspace{0.20\hsize}
\includegraphics[angle=-90,width=0.80\hsize]
 {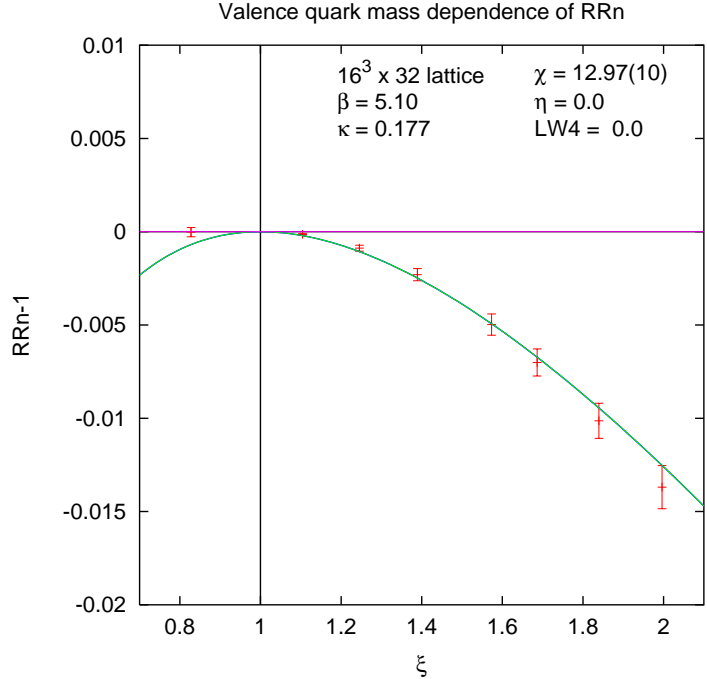}
\end{minipage}
\end{center}
\vspace*{-2em}
\begin{center}
\parbox{0.8\linewidth}{\caption{\label{fig05}\em
 Numerical results of the qq+q Collaboration on $16^3 \cdot 32$ lattice
 at ($\beta=5.1,\,\kappa=0.177$): one parameter fit of
 $(RRn-1)=\chi_S(1-\xi+\log\xi)/(32\pi^2)$ (``pure chiral log``).}}
\end{center}
\vspace*{-1em}
\end{figure}

 The result of this first study was that the formulas like
 (\ref{eq4.118})-(\ref{eq4.120}) describe well the dependence on both
 sea and valence quark masses, in particular if some generic NNLO
 terms are included.
 As an example of the fits see Figure~\ref{fig05}.
 First crude estimates of L-G constants, renormalized at scale
 $f_0\sqrt{\chi_R}$, gave with $\chi_R = 33.5(2.4)$
\be\label{eq4.123}
L_{R5} = 3.00(19) \cdot 10^{-3} \ , \hspace{3em}
(2L_{R8}-L_{R5}) = -6.25(52) \cdot 10^{-4} \ .
\ee
 From the sea quark mass dependence it was obtained
\bea\label{eq4.124}
(2L_{R4}+L_{R5}) &=& 4.34(28) \cdot 10^{-3} \ ,
\nonumber \\
(4L_{R6}+2L_{R8}-2L_{R4}-L_{R5}) &=& -9.1(6.4) \cdot 10^{-5} \ ,
\nonumber \\
\frac{\Lambda_3}{f_0} &=& 6.51(57) \ ,
\nonumber \\
\frac{\Lambda_4}{f_0} &=& 22.9(1.5)
\eea
 These numnbers can only be taken as crude estimates, because they
 come from a point with coarse lattice spacing and no continuum
 extrapolation has been performed.

\section{Outlook}\label{sec5}

 The present goal of numerical Monte Carlo investigations is to perform
 dynamical quark simulations with light quarks in large volumes.
 After about twenty-thirty years of hard work -- which can be considered
 as the preparation -- the presently available computer resources and
 algorithmic developments make this goal achievable.
 The big question is, can we validate QCD as the true theory of strong
 interactions by comparing the results with experimental knowledge?
 After this will be done, lattice gauge theorists will be able to extend
 their research area to study at the non-perturbative level a broader
 class of Quantum Field Theories not just QCD.

\subsection{Beyond QCD}\label{sec5.1}

 The further development of lattice regularized Quantum Field Theories
 will reflect how the two basic theoretical problems of the Electroweak
 Standard Model will be resolved in a {\em ``beyond the Standard Model''}
 framework.
 These two problems are:
\begin{itemize}
\item
 The {\em triviality} of the Higgs-Yukawa sector: as a consequence of
 appearance of Landau-Pomeranchuk poles there are cut-off dependent
 upper bounds on the Higgs- and Yukawa-couplings, which tend to zero
 for infinite cut-off (i.e.~zero lattice spacing).
\item
 It is very difficult to define {\em chiral gauge theories} in lattice
 regularization -- although they are required for the electroweak
 sector.
 Mirror fermion states with opposite chirality appear and it is
 difficult to separate the mass scale of the mirror fermion sector from
 the known chiral sector \cite{Mirror}.
 By including the mirror fermion sector the theory becomes vector-like
 (non-chiral).
\end{itemize}

 These problems become acute at the TeV scale and need some solution
 in a near future -- in particular based on the experimental input
 expected from LHC.
 There are several ways how these problems could perhaps be solved:
\begin{enumerate}
\item
 {\em Supersymmetric} extensions of the Standard Model: the improvement
 of the divergence structure due to supersymmetry (the solution of the
 ``hierarchy problem'' because of the absence of quadratic divergences)
 may solve both of the above problems.
 The mirror states could perhaps be shifted to the grand unification
 scale.
\item
 {Technicolor}-type models based on some appropriate generalization
 of QCD may produce the low-energy chiral spectrum as bound states.
 The mirror fermions could be at the technicolor scale.
\item
 {\em Beyond QFT} models where more dimensions beyond four appear
 and/or quantum gravity effects play an important role already near
 the TeV scale.
\end{enumerate}

 Which one (if any) of these ways is realized in Nature is a very
 exciting question and will hopefully become clear in the not very
 far future.
 If possibility 1. is realized then lattice field theorists will
 have to work more on (at least partly) supersymmetric non-perturbative
 regularization schemes.
 The case of possibility 2. seems to be a more or less straightforward
 generalization of QCD.
 In case of 3. one probably has to abandon the traditional QFT
 framework and look for radically new approaches.

\vspace*{1em}
\noindent
{\large\bf Acknowledgments}

\noindent
 It is a pleasure to thank the organizers, the lecturers and the students
 of the Spring School on High Energy Physics in Jaca, Spain for the
 lively and inspiring atmosphere at the School.




\newpage
\begin{thebibliography}{99}
%
\bibitem{MM}
 I.~Montvay, G.~M\"unster,
 {\em Quantum Fields on a Lattice,} Cambridge University Press,
 1994.
%
\bibitem{Wilson}
 K.G.~Wilson,
 Phys.\ Rev.\  D {\bf 10} (1974) 2445.
%
\bibitem{WilsonKogut}
 K.~G.~Wilson and J.~B.~Kogut,
 Phys.\ Rept.\  {\bf 12} (1974) 75.
%
\bibitem{Symanzik}
 K.~Symanzik,
 Nucl.\ Phys.\  B {\bf 226} (1983) 187.
%
\bibitem{WeiszWohlert}
 P.~Weisz,
 Nucl.\ Phys.\ B {\bf 212} (1983) 1.
\\
 P.~Weisz and R.~Wohlert,
 Nucl.\ Phys.\ B {\bf 236} (1984) 397
 [Erratum-ibid.\ B {\bf 247} (1984) 544].
%
\bibitem{MILC}
 Fermilab Lattice, MILC and HPQCD Collaboration, A.S.~Kronfeld et al.,
 PoS LAT2005 (2005) 206,
 Int.\ J.\ Mod.\ Phys.\ {\bf A21} (2006) 713; hep-lat/0509169.
%
\bibitem{Frezzotti:tmqcd}
 R.~Frezzotti, P.~A.~Grassi, S.~Sint and P.~Weisz,
 Nucl.\ Phys.\ Proc.\ Suppl.\  {\bf 83} (2000) 941; hep-lat/9909003.
%
\bibitem{FrezzottiRossi:split}
 R.~Frezzotti and G.C.~Rossi,
 Nucl.\ Phys.\ Proc.\ Suppl.\  {\bf 128} (2004) 193; hep-lat/0311008.
%
\bibitem{FrezzottiRossi:Oa}
 R.~Frezzotti and G.C.~Rossi,
 JHEP {\bf 0408} (2004) 007; hep-lat/0306014;
%
\bibitem{Kaplan}
 D.~B.~Kaplan,
 Phys.\ Lett.\  B {\bf 288} (1992) 342; hep-lat/9206013.
%
\bibitem{NarayananNeuberger}
 R.~Narayanan and H.~Neuberger,
 Phys.\ Lett.\  B {\bf 302} (1993) 62; hep-lat/9212019.
%
\bibitem{Shamir}
 Y.~Shamir,
 Nucl.\ Phys.\  B {\bf 406} (1993) 90; hep-lat/9303005.
%
\bibitem{GinspargWilson}
 P.~H.~Ginsparg and K.~G.~Wilson,
 Phys.\ Rev.\  D {\bf 25} (1982) 2649.
%
\bibitem{Luscher:chiral}
 M.~Luscher,
 Phys.\ Lett.\  B {\bf 428} (1998) 342; hep-lat/9802011.
%
\bibitem{Hasenfratz}
 P.~Hasenfratz,
 Nucl.\ Phys.\ Proc.\ Suppl.\  {\bf 63} (1998) 53; hep-lat/9709110.
%
\bibitem{NeubergerKikukawa}
 Y.~Kikukawa, H.~Neuberger and A.~Yamada,
 Nucl.\ Phys.\  B {\bf 526} (1998) 572; hep-lat/9712022.
%
\bibitem{HasenfratzLalienaNiedermayer}
 P.~Hasenfratz, V.~Laliena and F.~Niedermayer,
 Phys.\ Lett.\  B {\bf 427} (1998) 125; hep-lat/9801021.
%
\bibitem{NielsenNinomiya}
 H.~B.~Nielsen and M.~Ninomiya,
 Nucl.\ Phys.\  B {\bf 185} (1981) 20
 [Erratum-ibid.\  B {\bf 195} (1982) 541]. \\
 H.~B.~Nielsen and M.~Ninomiya,
 Nucl.\ Phys.\  B {\bf 193} (1981) 173.
%
\bibitem{HernandezJansenLuscher}
 P.~Hernandez, K.~Jansen and M.~Luscher,
 Nucl.\ Phys.\  B {\bf 552} (1999) 363; hep-lat/9808010.
%
\bibitem{GoltermanShamir:aoki}
 M.~Golterman and Y.~Shamir,
 Phys.\ Rev.\  D {\bf 68} (2003) 074501; hep-lat/0306002.
%
\bibitem{Morningstar}
 C.~Morningstar,
 hep-lat/0702020.
%
\bibitem{Metropolis}
 N.~Metropolis, A.W.~Rosenbluth, M.N.~Rosenbluth, A.H.~Teller
 and E.~Teller,
 J.\ Chem.\ Phys., {bf 21} (1953) 1087,
%
\bibitem{PetcherWeingarten}
 D.~H.~Weingarten and D.~N.~Petcher,
 Phys.\ Lett.\  B {\bf 99} (1981) 333.
%
\bibitem{HMC}
 S.~Duane, A.D.~Kennedy, B.J.~Pendleton, D.~Roweth,
 Phys.\ Lett.\ {\bf B195} (1987) 216.
%
\bibitem{PHMC}
 R.~Frezzotti and K.~Jansen,
 Phys.\ Lett.\  B {\bf 402} (1997) 328; hep-lat/9702016.\\
 R.~Frezzotti and K.~Jansen,
 Nucl.\ Phys.\  B {\bf 555} (1999) 395; hep-lat/9808011.\\
 R.~Frezzotti and K.~Jansen,
 Nucl.\ Phys.\  B {\bf 555} (1999) 432; hep-lat/9808038.
%
\bibitem{MontvayScholz}
 I.~Montvay and E.~Scholz,
 Phys.\ Lett.\  B {\bf 623} (2005) 73; hep-lat/0506006.\\
 E.~E.~Scholz and I.~Montvay,
 PoS {\bf LAT2006} (2006) 037; hep-lat/0609042.
%
\bibitem{RHMC}
 M.~A.~Clark and A.~D.~Kennedy,
 Phys.\ Rev.\ Lett.\  {\bf 98} (2007) 051601; hep-lat/0608015.
%
\bibitem{FrezzottiJansen}
 R.~Frezzotti and K.~Jansen,
 Nucl.\ Phys.\ Proc.\ Suppl.\  {\bf 63} (1998) 943; hep-lat/9709033.
%
\bibitem{Weinberg}
 S.~Weinberg,
 Physica A {\bf 96} (1979) 327.
%
\bibitem{GasserLeutwyler}
 J.~Gasser and H.~Leutwyler,
 Annals Phys.\  {\bf 158} (1984) 142.
%
\bibitem{BernardGolterman}
 C.~W.~Bernard and M.~F.~L.~Golterman,
 Phys.\ Rev.\  D {\bf 49} (1994) 486; hep-lat/9306005.
%
\bibitem{ETMC}
 Ph.~Boucaud {\it et al.}  [ETM Collaboration],
 hep-lat/0701012.
%
\bibitem{SharpeSingleton}
 S.~R.~Sharpe and R.~L.~.~Singleton,
 Phys.\ Rev.\  D {\bf 58} (1998) 074501; hep-lat/9804028.
%
\bibitem{RupakShoresh}
 G.~Rupak and N.~Shoresh,
 Phys.\ Rev.\  D {\bf 66} (2002) 054503; hep-lat/0201019.
%
\bibitem{qq+q}
 F.~Farchioni, C.~Gebert, I.~Montvay, E.~Scholz and L.~Scorzato  [qq+q
 Collaboration],
 Phys.\ Lett.\  B {\bf 561} (2003) 102; hep-lat/0302011. \\
 F.~Farchioni, I.~Montvay, E.~Scholz and L.~Scorzato  [qq+q Collaboration],
 Eur.\ Phys.\ J.\  C {\bf 31} (2003) 227; hep-lat/0307002. \\
 F.~Farchioni, I.~Montvay and E.~Scholz  [qq+q Collaboration],
 Eur.\ Phys.\ J.\  C {\bf 37} (2004) 197; hep-lat/0403014.
%
\bibitem{Mirror}
 I.~Montvay,
 Phys.\ Lett.\  B {\bf 199} (1987) 89.
%
\end{thebibliography}
\end{document}